\documentclass[aps,prc,showpacs,showkeys,nofootinbib]{revtex4}

\usepackage[dvips]{graphicx}
\usepackage{dcolumn}
\usepackage{amsthm}
\usepackage{bm}

\newtheorem{dfn}{Definition}

\begin{document}

\title{Quantum dynamics of evolution of flat universe in the first stage}



\author{S.~P.~Maydanyuk\thanks{\emph{E-mail:} maidan@kinr.kiev.ua}}

\affiliation{Institute for Nuclear Research, National Academy of Science of Ukraine, Kiev, 03680, Ukraine}

\date{\small\today}

\begin{abstract}
Process of formation of the universe with its further expansion in the first evolution stage is investigated in the framework of Friedmann-Robertson-Walker metrics on the basis of quantum model, where a new type of matter is introduced, which energy density is dependent on velocity of the expansion.
It is shown that such an improvement of the model forms potential barrier for the flat universe at $k=0$ (in contrast with generalized Chaplygin gas model).
Peculiarities of wave function are analyzed in details, which is calculated by fully quantum (non-semiclassic) approach, for the different barrier regions and stages of evolution.
Resonant influence of the initial and boundary conditions on the barrier penetrability is shown (in contrast with Vilenkin and Hawking approaches).
In order to perform a comparative analysis, how much quickly the universe is expanded by different models, new quantum definitions of velocity and Hubble function are introduced.
These notions allow us to study dynamics of evolution of universe in quantum cosmology both in the first stage, and in later times.
\end{abstract}



\maketitle

\section{Introduction
\label{sec.introduction}}

Data of astronomical observations suggest on speeding up character of expansion of the present universe. This is resent observations of supernova of type Ia (SNe Ia) \cite{Riess.1998.AJ,Perlmutter.1999.AJ,Tonry.2003.AJ}, more recent cosmic microwave background radiation (CMBR) data~\cite{Spergel.2003.AJS,Bennett.2003.AJS,Tegmark.2004.PRD}, clusters of galaxies~\cite{Pope.1999.AJ}, etc.
This has been designed as ``dark energy'' effect \cite{Sahni.2002.CQG}.
In order to explain this phenomenon, different approaches in cosmology have been intensively developed. One can divide them into two groups.
To the first group we can include models, describing evolution of expansion of the universe at present time. However, these models mainly are not quantum, and they do not describe the initial stage in details.
One of the characteristics used in these models is the parameter of Hubble: from comparison of its calculated values with observational data some conclusions are made about the restrictions applied on the considered models (for example, see~\cite{Lu.2009.EPJC}) and, sometimes, information about character of the evolution of the universe at present times is extracted.
To the second group one can include the models oriented on study of the formation of the universe and its further evolution in the first stage (i.e., Big Bang). Here, start of expanding of the universe is associated with tunneling transition through the barrier, usually based on quantum cosmology~\cite{AcacioDeBarros.2007.PRD,Monerat.2007.PRD}.
In particular, string and brane models~\cite{Brandenberger.1989.NPB,Park.2000.PRD,Gasperini.2003.PR,%
Battefeld.2006.RMP,Veneziano.v265.PLB,Cline.2006.hep-th.0612129}, and also multidimensional models~\cite{Carugno.1996.PRD} are intensively investigated in this group.
However, in frameworks of such models it is usually very difficult to hold out the calculations to the present times, and to connect them with data of astronomical observations.
So, a bridge between these two research directions is needed.

The quantum description of the formation of the universe and its evolution in the first stage is a particular difficult problem. From the literature we find that the expanded number of papers with practical calculations was performed in frameworks of the semiclassical approximation (for example, see~\cite{Vilenkin.1988.PRD,Vilenkin.1995,Casadio.2005.PRD.D71,Casadio.2005.PRD.D72,Luzzi.2006.PhD}).
However, for more accurate study one would like to renounce such approximations, and, sooner or later, it will have to do it.
Sometimes researchers use exactly solvable potentials (for example, see~\cite{Vilenkin.1988.PRD,Yurov.2001.CQG,Yurov.2004.TMP,Yurov.2005.PRD,Yurov.2009.TMP,Yurov.2011.TMP,Garcia.2006.IJTP}).
But the total number of such initial potentials can be counted on the fingers, while proper grounds for such research should be tools for work with barriers of arbitrarily shape and without semiclassical approximations.
This has caused interest to development of methods of quantum mechanics specially oriented on quantum cosmology (for example, see~\cite{Maydanyuk.2003.PhD-thesis,Rubakov.2002.PRD,Esposito.2012.JPA}).

One can understand the most clearly the quantum properties of the formation of the universe in the Friedmann-Robertson-Walker (FRW) model. For example, in~\cite{Maydanyuk.2011.EPJP} the resonant behavior of the penetrability of the barrier in dependence on the chosen initial and boundary conditions (i.e. coordinate of the start of wave, position of the external boundary outside the barrier, energy of the radiation) was opened.
This result can strongly (up to several thousand of percents) change the results obtained by other authors in the semiclassical approximation.
Another result of this paper (and \cite{Maydanyuk.2008.EPJC,Maydanyuk.2010.IJMPD}) is that space-time in the first stage of evolution of the universe is discrete rather than continuous. But, in later times this property gradually decreases and is lost. At the same time, the semiclassical methods are not sensitive to such peculiarity.

However, as mentioned in \cite{Beckwith.2012.JMP}, in such an approach the barrier is not formed for the flat universe at $k = 0$, and other theoretical grounds should be constructed than quantum mechanics provides this. In this paper, we introduce a new type of matter, the energy density of which depends on the velocity of expansion of the universe. We show that the inclusion of such a type of matter forms the potential barrier, and such an approach would be an some alternative to generalized Chaplygin gas model~\cite{Chaplygin.1904,Kamenshchik.2001.PLB}. For a comparative analysis, how fast the universe is expanding in frameworks of the different models, we introduce new quantum definitions of the velocity of expansion of the universe and the function of Hubble. This basis allows us to investigate dynamics in quantum cosmology, with hope to become a basis for testing the quantum models via astronomical observations.

\section{Formulation of quantum dynamics of evolution of universe in Friedmann-Robertson-Walker metric
\label{sec.model}}

\subsection{Model with energy density dependent on velocity of expansion
\label{sec.model.1}}

We shall start from consideration of FRW model.
We write action in the form (for example, see ~\cite{Vilenkin.1995}, Eq.(1), p.~2):
\begin{equation}
\begin{array}{cc}
  S = \displaystyle\int \sqrt{-g}\: \biggl( \displaystyle\frac{R}{16\pi\,G} - \rho \biggr)\; dx^{4}, &
  R =
  \displaystyle\frac{6\dot{a}^{2} + 6a\ddot{a} + 6k}{a^{2}},
\end{array}
\label{eq.model.1.1}
\end{equation}
where $R$ is Ricci scalar,
$a$ is the scale factor,
an overdot denotes a derivative with respect on cosmic time $t$,
$k$ is curvature of spatial section which equals to $+1$, $0$ or $-1$ (for open, flat and closed universe, correspondingly),
$G$ is Newtonian constant.
The energy density in presence of radiation and dark sector can be presented in the form:
\begin{equation}
\begin{array}{cc}
  \rho\,(a) = \rho_{\rm gCg}\,(a) + \displaystyle\frac{\rho_{\rm rad}}{a^{4}}, &
  \rho_{\rm gCg}\,(a) = \biggl( A + \displaystyle\frac{B}{a^{3\,(1+\alpha)}} \biggr)^{1/(1+\alpha)}.
\end{array}
\label{eq.model.1.2}
\end{equation}
Here, the first item $\rho_{\rm gCg}\,(a)$ represents the generalized Chaplygin gas, including dark energy term $A$ and dark matter term $B$, the second term $\rho_{\rm rad}$ describes energy density of radiation.

But, from literature we find that practically all calculations of quantum rates of universe evolution in the first stage have been performed for the case of $k=1$ only (for example, see~\cite{AcacioDeBarros.2007.PRD,Monerat.2007.PRD}). In such a case, potential used in the Wheeler-De Witt equation has a barrier, and formation of universe is described as tunneling transition through it.
At the same time, as it was indicated in \cite{Beckwith.2012.JMP}, for $k=0$ different scenarios of generalized Chaplygin gas model do not give the barrier, and this causes questions to applicability of the quantum description of formation of universe and its evolution, in general.
By such a reason, for such a special case of the flat universe \textit{at initial} (i.e. for $k=0$) we shall investigate a new scenario of evolution of the universe where we include a new component of energy density which is dependent not only on the scale factor $a$, but also on its derivative $\dot{a}$, i.e. velocity of expansion.

\vspace{3mm}
\noindent
\underline{Hypothesis:}
\emph{We shall introduce into the model the new form of energy (matter) of unknown nature which density is dependent on $\dot{a}$ as}
\begin{equation}
  \rho_{\rm vel}(a, \dot{a}) = \rho_{0, \rm vel}(a) \; \dot{a}^{n},
\label{eq.model.1.3}
\end{equation}
\emph{where $n$ is some constant parameter. We assume that such a behavior of density can be caused by presence dissipative (self-)\,interacting forces or forces of inversed form (i.e. quantum sources) in this type of matter. In particular, we assume that such a type of matter could be some variant of realization of dark energy/matter, and could be connected with origin of curvature of the initially flat space-time.}
In this paper, we shall investigate if inclusion of such a type of matter allows us to form barrier for flat universe which produces non-unite penetrability with non-zero reflection.






\vspace{3mm}
Now let us compose the total energy density. Here, we include dark matter and dark energy terms separated explicitly, term of radiation and the new density~(\ref{eq.model.1.3}):
\begin{equation}
\begin{array}{cc}
  \rho\,(a, \dot{a}) =
    \rho_{\rm vac} +
    \displaystyle\frac{\rho_{\rm dust}}{a^{3}(t)} +
    \displaystyle\frac{\rho_{\rm rad}}{a^{4}(t)} +
    \rho_{0, \rm vel}(a) \; \dot{a}^{n} =
  \rho_{1}\,(a) + \rho_{0, \rm vel}(a) \; \dot{a}^{n}.
\end{array}
\label{eq.model.1.4}
\end{equation}
Here, $\rho_{\rm vac}$ is energy density of vacuum which can be connected with cosmological constant as $\rho_{\rm vac} = \Lambda / (8\pi G)$ and represents dark energy term,
$\rho_{\rm dust}$ is constant factor of dark matter in dust-like form.
Substituting (\ref{eq.model.1.4}) to (\ref{eq.model.1.1}), we obtain lagrangian:
\begin{equation}
  L\,(a,\dot{a}) =
  \mathcal{N}\;\displaystyle\frac{3\,a}{8\pi\,G}\:
  \biggl(-\dot{a}^{2} + k -
    \displaystyle\frac{8\pi\,G}{3}\; a^{2}\, \bigl[\rho(a) + \rho_{0, \rm vel} (a)\, \dot{a}^{n} \bigr] \biggr).
\label{eq.model.1.5}
\end{equation}
In particular, for flat universe of closed type (i.e. at $\rho = \rho_{\rm vac}$, $\alpha=0$, $k=1$, $\mathcal{N}=1$, $\rho_{0,\rm vel}=0$) we have:
\begin{equation}
\begin{array}{ccl}
  L\,(a,\dot{a}) =
  \displaystyle\frac{3}{4\pi\,G}\cdot
  \displaystyle\frac{a}{2}\cdot
  \biggl[-\dot{a}^{2} + k -
  \displaystyle\frac{8\pi\,G}{3}\:\rho_{\rm vac}\,a^{2} \biggr] \sim
  \displaystyle\frac{a}{2}\: \Bigl(1 -  \dot{a}^{2} - H_{\rm vac}^{2}\,a^{2} \Bigr), &
  H_{\rm vac} = \sqrt{\displaystyle\frac{\Lambda}{3}},
\end{array}
\label{eq.model.1.6}
\end{equation}
that coincides with eq.~(11) in~\cite{Vilenkin.1995} (up to normalized factor, see p.~4 in that paper).
Defining hamiltonian as
\begin{equation}
\begin{array}{ccl}
\vspace{1mm}
  h\,(a,p_{a}) & = & p\,\dot{a} - L\,(a,\dot{a}) =
  \mathcal{N}\;a\;
  \Bigl\{
   - \displaystyle\frac{3}{8\pi\,G}\:
    \bigl[\dot{a}^{2} + k \bigr] +
    a^{2}\,\rho(a) +
    (1-n)\; a^{2}\,\rho_{0,\rm vel}(a)\; \dot{a}^{n}
  \Bigr\},
\end{array}
\label{eq.model.1.7}
\end{equation}
where $p$ is momentum conjugated to generalized coordinate $a$, we find:
\begin{equation}
\begin{array}{cl}
\vspace{1mm}
\mbox{at $n=1$ : } &
  h\,(a,p_{a}) =
  - \displaystyle\frac{2\pi\,G}{3\mathcal{N}\, a}\;
      \biggl\{ p_{a}^{2} +
      2\; p_{a}\;\mathcal{N}\; a^{3}\; \rho_{0, \rm vel}(a) +
      \mathcal{N}^{2}\; a^{6}\; \rho_{0, \rm vel}^{2}(a) +
      \Bigl( \displaystyle\frac{3\,\mathcal{N}}{4\pi\,G} \Bigl)^{2}\,k\, a^{2} -
      \displaystyle\frac{3\,\mathcal{N}^{2}}{2\pi\,G}\;
    a^{4}\,\rho(a)
  \biggr\}, \\

\mbox{at $n=2$ : } &
  h\,(a,p_{a}) =
  -\,\displaystyle\frac{1}{4\,\mathcal{N}\,a\, \Bigl[ a^{2}\; \rho_{0, \rm vel}(a) + 3/(8\pi\,G) \Bigr]}\;
    \biggl\{
      p_{a}^{2}\; + \;
      4\,\mathcal{N}^{2}\;a^{2}\, \Bigl[ a^{2}\; \rho_{0, \rm vel}(a) + 3/(8\pi\,G) \Bigr]\: \times \\
  & \times \;
    \Bigl( \displaystyle\frac{3\,k}{8\pi\,G} - a^{2}\,\rho(a) \Bigr)
  \biggr\}.
\end{array}
\label{eq.model.1.8}
\end{equation}
In particular, hamiltonian of our previous model \cite{Maydanyuk.2011.EPJP} (i.e. without density $\rho_{\rm vel}$) can be obtained at $\rho_{0, \rm vel} \to 0$.
Here, from each expression in (\ref{eq.model.1.8}) we obtain:
\begin{equation}
\begin{array}{cll}
  h\,(a,p_{a}) & = &
  - \displaystyle\frac{2\pi\,G}{3\mathcal{N}\, a}\;
      \biggl\{ p_{a}^{2} +
      \Bigl( \displaystyle\frac{3\,\mathcal{N}}{4\pi\,G} \Bigl)^{2}\,k\, a^{2} -
      \displaystyle\frac{3\,\mathcal{N}^{2}}{2\pi\,G}\;
    a^{4}\,\rho(a)
  \biggr\} = 
  -\:\displaystyle\frac{\mathcal{N}}{a}\;
  \biggl\{
    \displaystyle\frac{2\pi\,G}{3\,\mathcal{N}^{2}}\: p_{a}^{2} +
    a^{2}\,\displaystyle\frac{3\,k}{8\pi\,G} -
    a^{4}\,\rho(a) \biggr\}.
\end{array}
\label{eq.model.1.9}
\end{equation}

Now we apply quantization, after which we obtain stationary Wheeler-De Witt equation
(for example, see~\cite{Vilenkin.1995}, (16)--(17) p.~4; see~\cite{Wheeler.1968,DeWitt.1967,Rubakov.2002.PRD}).
After multiplication on corresponding factors we obtain:
\begin{equation}
\begin{array}{lcl}
\vspace{2mm}
  \mbox{\rm at $n=1$ : } &
    \biggl\{
      -\: \displaystyle\frac{\partial^{2}}{\partial a^{2}}\; - \;
      i\, 2\, \mathcal{N}\, a^{3}\; \rho_{0, \rm vel}(a)\: \displaystyle\frac{\partial}{\partial a}\; +
      \mathcal{N}^{2}\; a^{6}\; \rho_{0, \rm vel}^{2}(a) +
      \Bigl( \displaystyle\frac{3\,\mathcal{N}}{4\pi\,G} \Bigl)^{2}\,k\, a^{2} -
      \displaystyle\frac{3\,\mathcal{N}^{2}}{2\pi\,G}\;
    a^{4}\,\rho(a)
  \biggr\}\; \varphi(a) = 0, \\

\vspace{2mm}
  \mbox{\rm at $n=2$ : } &
    \biggl\{
      -\: \displaystyle\frac{\partial^{2}}{\partial a^{2}}\; + \;
      4\,\mathcal{N}^{2}\;a^{2}\, \Bigl[ a^{2}\; \rho_{0, \rm vel}(a) + 3/(8\pi\,G) \Bigr]\:
    \Bigl( \displaystyle\frac{3\,k}{8\pi\,G} - a^{2}\,\rho(a) \Bigr)
  \biggr\}\; \varphi(a) = 0, \\

  \mbox{\rm model \cite{Maydanyuk.2011.EPJP} : } &
  \biggl\{
    -\: \displaystyle\frac{\partial^{2}}{\partial a^{2}} +  V (a)
  \biggr\}\; \varphi(a) = 0,\quad

  V (a) = -\, \displaystyle\frac{12\,\mathcal{N}^{2}}{8\pi\,G}\;
    \Bigl[
      - \displaystyle\frac{3}{8\pi\,G}\: k\,a^{2} +
      \rho_{\rm rad} +
      \rho_{\rm gCg}\, a^{4}
    \Bigr].
\end{array}
\label{eq.model.1.10}
\end{equation}

\subsection{Case $n=2$ at $k=0$
\label{sec.model.2}}

Further in this paper, we shall be interesting in the model with density (\ref{eq.model.1.4}) at $k=0$ and $n=2$.
For the second equation in (\ref{eq.model.1.10}) we have:
\begin{equation}
\begin{array}{ccl}
\vspace{1mm}
  V\,(a) & = &
    -\,12\,\mathcal{N}^{2}\;
    \Bigl[
      \displaystyle\frac{a^{2}\; \rho_{0, \rm vel}(a)}{3} +
      \displaystyle\frac{1}{8\pi\,G}
    \Bigr]\:
    \Bigl(
      \rho_{\rm dust}\,a +
      \rho_{\rm rad} +
      \rho_{\rm vac}\, a^{4}
    \Bigr).
\end{array}
\label{eq.model.2.1}
\end{equation}

\subsubsection{Energy density $\rho_{\rm vel}$ at large $a$: vacuum energy vs new type of energy.
\label{sec.model.2.1}}


In particular, at large $a$ potential (\ref{eq.model.2.1}) without vacuum energy density $\rho_{\rm vac}$ obtains the same behavior as potential of model \cite{Maydanyuk.2011.EPJP} with this vacuum energy term. Comparing potential~(\ref{eq.model.2.1}) with the third equation in (\ref{eq.model.1.10}), we obtain:
\begin{equation}
\begin{array}{cclcc}
  \biggl(
    V\,(a) =
    -\,\displaystyle\frac{3\,\mathcal{N}^{2}}{2\pi\,G}\; \rho_{\rm vac}\, a^{4} =
    -\,4\,\mathcal{N}^{2}\; a^{2}\; \rho_{0, \rm vel}(a)\: \rho_{\rm dust}\,a
  \Bigr) & \to &
  \biggl(
    \rho_{0,\rm vel}(a) =
    \displaystyle\frac{3}{8\pi\,G}\, \displaystyle\frac{\rho_{\rm vac}}{\rho_{\rm dust}}\; a
  \Bigr)
\end{array}
\label{eq.model.2.1.1}
\end{equation}
or
\begin{equation}
\begin{array}{cclcc}
  \rho_{0,\rm vel}(a) = c_{\rm vel} \cdot a, &
  c_{\rm vel} = \displaystyle\frac{3}{8\pi\,G}\, \displaystyle\frac{\rho_{\rm vac}}{\rho_{\rm dust}}.
\end{array}
\label{eq.model.2.1.2}
\end{equation}
So, at choice of $\rho_{0,\rm vel}(a)$ in form (\ref{eq.model.2.1.2}), the potential (\ref{eq.model.2.1}) without vacuum energy density has the same shape at large $a$ as potential of the model~\cite{Maydanyuk.2011.EPJP}.

\subsubsection{Energy density $\rho_{\rm vel}$ at small $a$: formation of the barrier.
\label{sec.model.2.2}}


One can form barrier introducing density $\rho_{\rm 0, vel}$ in the form:
\begin{equation}
\begin{array}{llll}
  \mbox{\rm dust-like energy:} &
  \rho_{0, \rm vel}(a) = \rho_{0}/ a^{3}&
  \mbox{ or} &
  \rho_{\rm vel}(a, \dot{a}) = \rho_{0}\, \dot{a}^{2} / a^{3}, \\

  \mbox{\rm radiation-like energy:} &
    \rho_{0, \rm vel}(a) = \rho_{0}/ a^{4} &
  \mbox{ or} &
  \rho_{\rm vel}(a, \dot{a}) = \rho_{0}\, \dot{a}^{2} / a^{4}, \\

  \mbox{\rm other types of matter:} &
    \rho_{0, \rm vel}(a) = \rho_{0}/ a^{m} &
  \mbox{ or} &
  \rho_{\rm vel}(a, \dot{a}) = \rho_{0}\, \dot{a}^{2} / a^{m},
  \quad m>2.
\end{array}
\label{eq.model.2.2.1}
\end{equation}
From here one can write a generalized form of energy density of dust-like matter or radiation, which forms barrier at small $a$:
\begin{equation}
\begin{array}{llll}
  \rho_{\rm dust}^{gen}(a, \dot{a}) =
  \rho_{\rm dust}\, \bigl[1 + c_{\rm dust} \cdot \dot{a}^{2} \bigr]/ a^{3} &
  \mbox{\rm or } &
  \rho_{\rm rad}^{gen}(a, \dot{a}) =
   \rho_{\rm rad}\, \bigl[1 + c_{\rm rad} \cdot \dot{a}^{2} \bigr]/ a^{4}.
\end{array}
\label{eq.model.2.2.2}
\end{equation}
Now we compose density $\rho_{0, \rm vel}$, adding its behavior (\ref{eq.model.2.1.2}) at large $a$:
\begin{equation}
\begin{array}{llll}
  \rho_{0, \rm vel}(a) = \rho_{\rm dust}\, c_{\rm dust}/a^{3} + c_{\rm vel} \cdot a &
  \mbox{\rm or } &
  \rho_{0, \rm vel}(a) = \rho_{\rm rad}\, c_{\rm rad}/a^{4} + c_{\rm vel} \cdot a.
\end{array}
\label{eq.model.2.2.3}
\end{equation}
In particular, using the first formula for a generalized dust-like matter, we write potential (\ref{eq.model.2.1}) (at $\mathcal{N}=1$, $8\pi\,G=1$, $\rho_{\rm vac}=0$):
\begin{equation}
\begin{array}{cclcc}
\vspace{1mm}
  V\,(a) & = &
    -12\, \Bigl\{ \rho_{\rm rad} + \rho_{\rm dust}\,a \Bigr\}\,
    \Bigl[ 1 + \rho_{0, \rm vel}(a) \cdot a^{2} /3 \Bigr]\; = \;
    -12\,
    \Bigl\{
      \displaystyle\frac{\rho_{\rm rad}\,\rho_{\rm dust}\,c_{\rm dust}}{3\,a}\; + \\
    & + &
      \Bigl[ \rho_{\rm rad} +
      \displaystyle\frac{\rho_{\rm dust}^{2}\,c_{\rm dust}}{3} \Bigr] +
      \rho_{\rm dust}\,a +
      \displaystyle\frac{\rho_{\rm rad}\, c_{\rm vel}}{3}\, a^{3} +
      \displaystyle\frac{\rho_{\rm dust}\, c_{\rm vel}}{3}\, a^{4}
    \Bigr\}.
\end{array}
\label{eq.model.2.2.4}
\end{equation}

\subsubsection{Comparison with Chaplygin gas model and choice of parameters
\label{sec.model.2.3}}

Some estimations of rates were performed on the basis of analysis of dynamic of wave packet tunneling through the potential barrier~\cite{Monerat.2007.PRD}. However, a case of $\alpha=1$ (i.e. for standard formulation of Chaplygin gas) corresponds to those calculations (see energy density (14) in that paper).
So, we shall use parameters of that paper for our calculations, in order to give other researchers more material for comparison and analysis.
According to that paper, let us rewrite energy density of the generalized Chaplygin gas in the following form:
\begin{equation}
\begin{array}{ccl}
  \rho_{\rm gCg}\, (a) \; = \;
  \biggl( A + \displaystyle\frac{B}{a^{3\,(1+\alpha)}} \biggr)^{1/(1+\alpha)} =
  \displaystyle\frac{1}{\pi}\,
  \biggl( \bar{A} + \displaystyle\frac{\bar{B}}{a^{3\,(1+\alpha)}} \biggr)^{1/(1+\alpha)},
\end{array}
\label{eq.model.2.3.1}
\end{equation}
where $\bar{A} = A\, \pi^{1+\alpha}$ and $\bar{B} = B\, \pi^{1+\alpha}$. We have:
\begin{equation}
\begin{array}{cllcc}
\vspace{1mm}
  1) & \mbox{\rm at small $a$:} &
  \rho_{\rm gCg}\,(a) \to
  \displaystyle\frac{B^{1/(1+\alpha)}}{a^{3}} \equiv \displaystyle\frac{\rho_{\rm dust}}{a^{3}}, \\
  2) & \mbox{\rm at large $a$:} &
  \rho_{\rm gCg} \to A^{1/(1+\alpha)} \equiv \rho_{\rm vac}.
\end{array}
\label{eq.model.2.3.2}
\end{equation}
In ~\cite{Monerat.2007.PRD} the following values $\bar{A}=0.001$, $\bar{B}=0.001$ were chosen (see Fig.~1 in that paper).
From here we find parameters:
\begin{equation}
\begin{array}{lcc}
\vspace{1mm}
  \rho_{\rm dust} & = &
    B^{1/(1+\alpha)} =
    \bar{B}^{1/(1+\alpha)}\, \pi^{-(1+\alpha) \cdot 1/(1+\alpha)} =
    \bar{B}^{1/(1+\alpha)} / \pi =
    0.001^{1/(1+\alpha)} / \pi, \\
  \rho_{\rm vac} & = &
    A^{1/(1+\alpha)} =
    \bar{A}^{1/(1+\alpha)}\, \pi^{-(1+\alpha) \cdot 1/(1+\alpha)} =
    \bar{A}^{1/(1+\alpha)} / \pi =
    0.001^{1/(1+\alpha)} / \pi.
\end{array}
\label{eq.model.2.3.3}
\end{equation}

\subsection{Operators of the function of Hubble and velocity of expansion of the universe
\label{sec.model.3}}

Let us consider how the velocity of the evolution of the universe can be defined in quantum approach. In classical mechanics the velocity of the particle is related to its momentum. In quantum mechanics, there is connection between the corresponding operators. According to basic positions of quantum mechanics, determination of the wave function $\Psi$ at some moment of time $t_{0}$, not only fully describes all quantum properties of the studied system at this time $t_{0}$, but also fully determines its evolution in all future times.
In other words, derivative $\partial \Psi / \partial t$ of the wave function on time at any given moment of time  $t_{0}$ is determined by this wave function $\Psi$ at $t_{0}$. If we want to use the principle of superposition, this dependence should be linear.
On this basis, instead of the stationary Wheeler-De Witt equation we shall be based on the following equation:
\begin{equation}
\begin{array}{ccc}
  i\hbar\: \displaystyle\frac{\partial \Psi}{\partial t} =
  \hat{h}\: \Psi,
\end{array}
\label{eq.model.3.1}
\end{equation}
where $\hat{h}$ is some unknown operator. Let us clarify what this operator should be to. If to assume that the universe expands classically (i.e. with high degree of confidence we can neglect by the quantum properties) at large values of the scale factor $a$,
then the wave function can be written as
\begin{equation}
  \Psi = A\, e^{iS / \hbar},
\label{eq.model.3.2}
\end{equation}
where $S$ is action. Substituting this expression for the wave function into (\ref{eq.model.3.1}) and neglecting by change of amplitude $A$ over time,
we find:
\begin{equation}
\begin{array}{ccc}
  \displaystyle\frac{\partial \Psi}{\partial t} =
  \displaystyle\frac{i}{\hbar}\,
  \displaystyle\frac{\partial S}{\partial t}\,
  \Psi,
\end{array}
\label{eq.model.3.3}
\end{equation}
Comparing this expression with definition (\ref{eq.model.3.1}), we conclude that in the limiting case the operator $\hat{h}$ is reduced to simple multiplication on value of $-\partial S / \partial t$, i.e. Hamiltonian function.
Now we have introduced time into quantum equation, connecting it with Hamiltonian operator and taking into account that the asymptotic representation of the wave function is connected with action as (\ref{eq.model.3.2}).

As the non-stationary quantum equation has already been defined, now we can define operator of the velocity using general rule of differentiation of operators over time 
(for example, see (19.1) in \cite{Landau.v3.1989}, p. 78--79).
\begin{dfn}[] 
\label{def.velocity}
We define operator of the velocity as
\begin{equation}
\begin{array}{ccc}
  \hat{\dot{a}} = \displaystyle\frac{i}{\hbar}\; (\hat{h}\, a - a\, \hat{h}).
\end{array}
\label{eq.model.3.4}
\end{equation}
\end{dfn}
\noindent
Here, we use requirement (as in standard quantum mechanics):
\begin{equation}
  \displaystyle\frac{\partial}{\partial t} \displaystyle\int |\Psi(a, t)|^{2}\; da = 0.
\label{eq.model.3.5}
\end{equation}
Substituting hamiltonian in form (\ref{eq.model.1.8}) or (\ref{eq.model.1.9}), we find:
\begin{equation}
\begin{array}{lll}
\vspace{2mm}
  \mbox{\rm at $n=1$ : } &
  \hat{\dot{a}} =
  \displaystyle\frac{i}{\hbar}\;
  \displaystyle\frac{8\pi G}{6 \mathcal{N} a}\;
  \biggl\{
    \displaystyle\frac{\partial}{\partial a}\; + \;
    i\, \mathcal{N}\, a^{3}\; \rho_{0, \rm vel}(a)
  \biggr\}, \\

\vspace{2mm}
  \mbox{\rm at $n=2$ : } &
  \hat{\dot{a}} =
  \displaystyle\frac{i}{\hbar}\,
  \displaystyle\frac{1}{2\,\mathcal{N}\,a\, \Bigl[ a^{2}\; \rho_{0, \rm vel}(a) + 3/(8\pi\,G) \Bigr]}\;
  \displaystyle\frac{\partial}{\partial a}, \\

  \mbox{\rm model~\cite{Maydanyuk.2011.EPJP}: } &
  \hat{\dot{a}} =
  \displaystyle\frac{i}{\hbar}\,
  \displaystyle\frac{8\pi G}{6\mathcal{N}\,a }\;
  \displaystyle\frac{\partial}{\partial a}.
\end{array}
\label{eq.model.3.6}
\end{equation}

\begin{dfn}[] 
\label{def.hubble}
We define operator of the function of Hubble as
\begin{equation}
  \hat{H}\,(a) = \displaystyle\frac{1}{a}\,\hat{\dot{a}}.
\label{eq.model.3.7}
\end{equation}
\end{dfn}
\noindent
According to this definition, we shall consider the parameter of Hubble at the given scale factor $a_{0}$, as action of certain operator $\hat{H}$ on the wave function at $a_{0}$. The wave function is not eigenfunction of operators of the velocity and the function of Hubble, as there are no any constant eigenvalues for these operators.
So, action of these operators on the wave function can be written as
\begin{equation}
\begin{array}{ll}
  \hat{\dot{a}}\: \Psi\,(a) = v(a)\, \Psi\,(a), &
  \hat{H}\,(a)\, \Psi\,(a) = H\,(a)\, \Psi\,(a).
\end{array}
\label{eq.model.3.8}
\end{equation}
Here, $v(a)$ and $H\,(a)$ are some functions changed in depending on $a$. Near to arbitrarily chosen value $a_{0}$ these functions tend to certain well-defined fixed values, which can be locally considered as eigenvalues of operators of the velocity and the function of Hubble at $a_{0}$. Thus, on the basis of the functions $v(a)$ and $H\,(a)$ we shall understand the velocity and the parameter of Hubble in the quantum approach.
For practical calculations, one can obtain these functions as
\begin{equation}
\begin{array}{lll}
  v\,(a) =
  \displaystyle\frac{\hat{\dot{a}}\; \Psi\,(a)}{\Psi\,(a)}, &
  H\,(a) =
  \displaystyle\frac{\hat{H}\; \Psi\,(a)}{\Psi\,(a)} =
  \displaystyle\frac{v\,(a)}{a}.
\end{array}
\label{eq.model.3.9}
\end{equation}

\subsection{Quantum definition of duration of existence of universe
\label{sec.model.4}}

As the function of Hubble has been defined, as next step we can define duration of existence of the universe. We shall be interesting in such a characteristic which is dependent on the scale factor $a$: this should allow to see clearly behavior of expansion (evolution) and compare such a dynamic for different model scenarios. So, we introduce the following characteristic:
\begin{equation}
\begin{array}{lll}
  t\,(a) =
  \displaystyle\int\limits_{a_{\rm min}}^{a}
  \displaystyle\frac{1} {\tilde{a}\, H\,(\tilde{a})}\; d\tilde{a}.
\end{array}
\label{eq.model.4.1}
\end{equation}
One can see that such a definition is analog of the classical definition of the universe age given in classical cosmology (for example, see~(36), p.~11 in~\cite{Trodden.TASI-2003}).

\subsection{Rescaling
\label{sec.model.5}}

Inside region $0<a<100$ potentials of the considered models above achieve essential values (that can cause serious difficulties in practical calculations and further analysis). But, they can be decreased via renormalization of the corresponding equations (\ref{eq.model.1.10}) (we shall call such a procedure as \emph{rescaling}).
By such a reason, let us pass to a new variable:
\begin{equation}
\begin{array}{cc}
  a_{\rm new} = \nu\, a_{\rm old}, &
  \nu = \sqrt{\displaystyle\frac{12\,\mathcal{N}^{2}}{8\pi\,G}}.
\end{array}
\label{eq.model.5.1}
\end{equation}
Now we obtain new equations for determination of the wave function:
\begin{equation}
\begin{array}{ccl}
\vspace{1mm}
  \mbox{\rm at $n=1$ : } &
    -\: \displaystyle\frac{\partial^{2}}{\partial a_{\rm new}^{2}}\; +
      \biggl( \displaystyle\frac{8\pi\,G}{12\,\mathcal{N}^{2}} \biggr) \cdot
      \biggl\{ - \:
      i\, 2\, \mathcal{N}\,\nu\, \Bigl(\displaystyle\frac{a_{\rm new}}{\nu}\Bigr)^{3}\;
        \rho_{0, \rm vel} \Bigl(\displaystyle\frac{a_{\rm new}}{\nu}\Bigr)\:
        \displaystyle\frac{\partial}{\partial a_{\rm new}}\; +
      \mathcal{N}^{2}\:
      \Bigl(\displaystyle\frac{a_{\rm new}}{\nu}\Bigr)^{6}\;
      \rho_{0, \rm vel}^{2} \Bigl(\displaystyle\frac{a_{\rm new}}{\nu}\Bigr)\; + \\
\vspace{3mm}
    &
      + \; \Bigl( \displaystyle\frac{3\,\mathcal{N}}{4\pi\,G} \Bigl)^{2}\,k\,
      \Bigl(\displaystyle\frac{a_{\rm new}}{\nu}\Bigr)^{2} -
      \displaystyle\frac{3\,\mathcal{N}^{2}}{2\pi\,G}\;
      \Bigl(\displaystyle\frac{a_{\rm new}}{\nu}\Bigr)^{4}\,
      \rho\Bigl(\displaystyle\frac{a_{\rm new}}{\nu}\Bigr)
  \biggr\}\; \varphi (a_{\rm new}) = 0, \\

\vspace{1mm}
  \mbox{\rm at $n=2$ : } &
      -\: \displaystyle\frac{\partial^{2}}{\partial a_{\rm new}^{2}}\; + \;
    8\pi\,G\,
    \Bigl[
      \displaystyle\frac{1}{3}\,
      \Bigl(\displaystyle\frac{a_{\rm new}}{\nu}\Bigr)^{2}\;
      \rho_{0, \rm vel}\Bigl(\displaystyle\frac{a_{\rm new}}{\nu}\Bigr) + 1/(8\pi\,G) \Bigr]\: \times\\
\vspace{3mm}
   & \times \;
    \Bigl[ \displaystyle\frac{3\,k}{8\pi\,G}\, \Bigl(\displaystyle\frac{a_{\rm new}}{\nu}\Bigr)^{2}\, -
      \Bigl(\displaystyle\frac{a_{\rm new}}{\nu}\Bigr)^{4}
      \rho\Bigl(\displaystyle\frac{a_{\rm new}}{\nu}\Bigr) \Bigr]\;
    \varphi(a_{\rm new}) = 0, \\

  \mbox{\rm model~\cite{Maydanyuk.2011.EPJP} : } &
  -\: \displaystyle\frac{\partial^{2}}{\partial a_{\rm new}^{2}}\; + \;
  \biggl\{
    \displaystyle\frac{3k}{8\pi\,G}\, \Bigl(\displaystyle\frac{a_{\rm new}}{\nu}\Bigr)^{2} -
    \Bigl(\displaystyle\frac{a_{\rm new}}{\nu}\Bigr)^{4}\,
      \rho\Bigl(\displaystyle\frac{a_{\rm new}}{\nu}\Bigr)
  \biggr\}\; \varphi(a_{\rm new}) = 0.
\end{array}
\label{eq.model.5.2}
\end{equation}
For operator of velocity (\ref{eq.model.3.6}), determined via new variable, we obtain:
\begin{equation}
\begin{array}{lll}
\vspace{2mm}
  \mbox{\rm at $n=1$ : } &
  \hat{\dot{a}} =
  \displaystyle\frac{i}{\hbar}\;
  \displaystyle\frac{8\pi G\, \nu^{2}}{6 \mathcal{N} a_{\rm new}}\;
  \biggl\{
    \displaystyle\frac{\partial}{\partial a_{\rm new}}\; + \;
    i\, \mathcal{N}\,\nu\, \Bigl(\displaystyle\frac{a_{\rm new}}{\nu}\Bigr)^{3}\;
    \rho_{0, \rm vel}\Bigl(\displaystyle\frac{a_{\rm new}}{\nu}\Bigr)
  \biggr\}, \\

\vspace{2mm}
  \mbox{\rm at $n=2$ : } &
  \hat{\dot{a}} =
  \displaystyle\frac{i}{\hbar}\,
  \displaystyle\frac{\nu^{2}}{2\,\mathcal{N}\,a_{\rm new}}\,
  \Bigl[ \Bigl(\displaystyle\frac{a_{\rm new}}{\nu}\Bigr)^{2}\:
    \rho_{0, \rm vel}\Bigl(\displaystyle\frac{a_{\rm new}}{\nu}\Bigr) + 3/(8\pi\,G) \Bigr]\;
  \displaystyle\frac{\partial}{\partial a_{\rm new}}, \\

   \mbox{\rm model~\cite{Maydanyuk.2011.EPJP}: } &
   \hat{\dot{a}} =
   \displaystyle\frac{i}{\hbar}\,
   \displaystyle\frac{8\pi G\, \nu^{2}}{6\mathcal{N}\,a_{\rm new}}\;
   \displaystyle\frac{\partial}{\partial a_{\rm new}}.
\end{array}
\label{eq.model.5.3}
\end{equation}



\section{The model~\cite{Maydanyuk.2011.EPJP}
\label{sec.results.2}}

\subsection{Tunneling deeply under the barrier
\label{sec.results.2.1}}

\begin{figure}[htbp]
\centerline{\includegraphics[width=65mm]{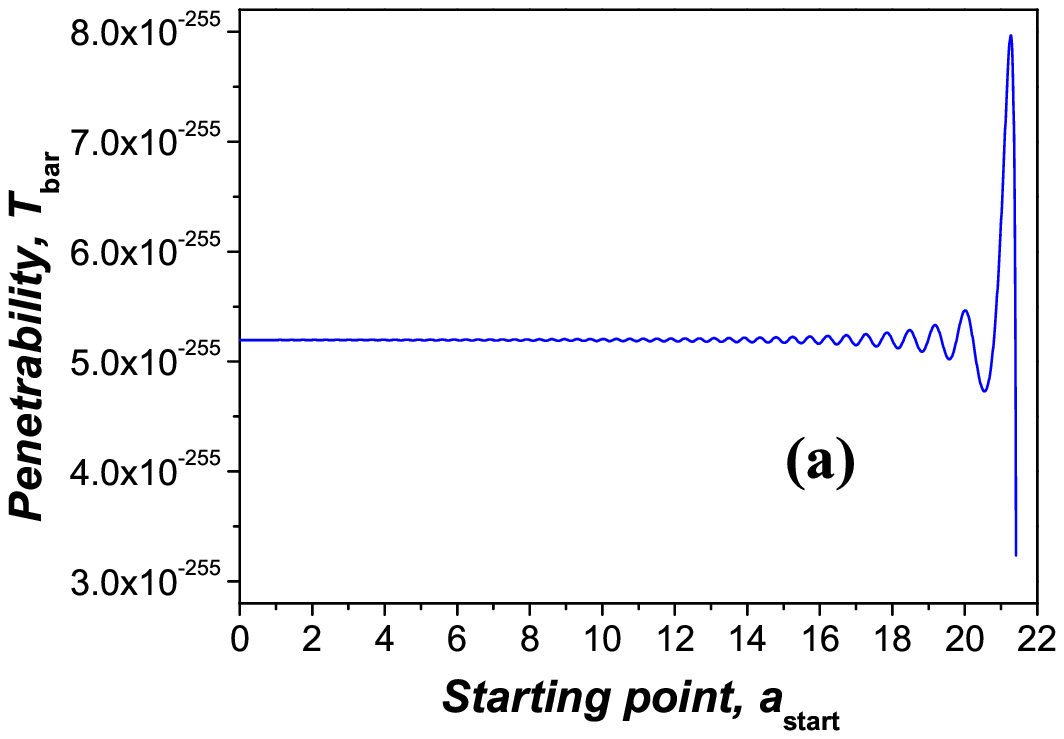}
\hspace{-7mm}\includegraphics[width=65mm]{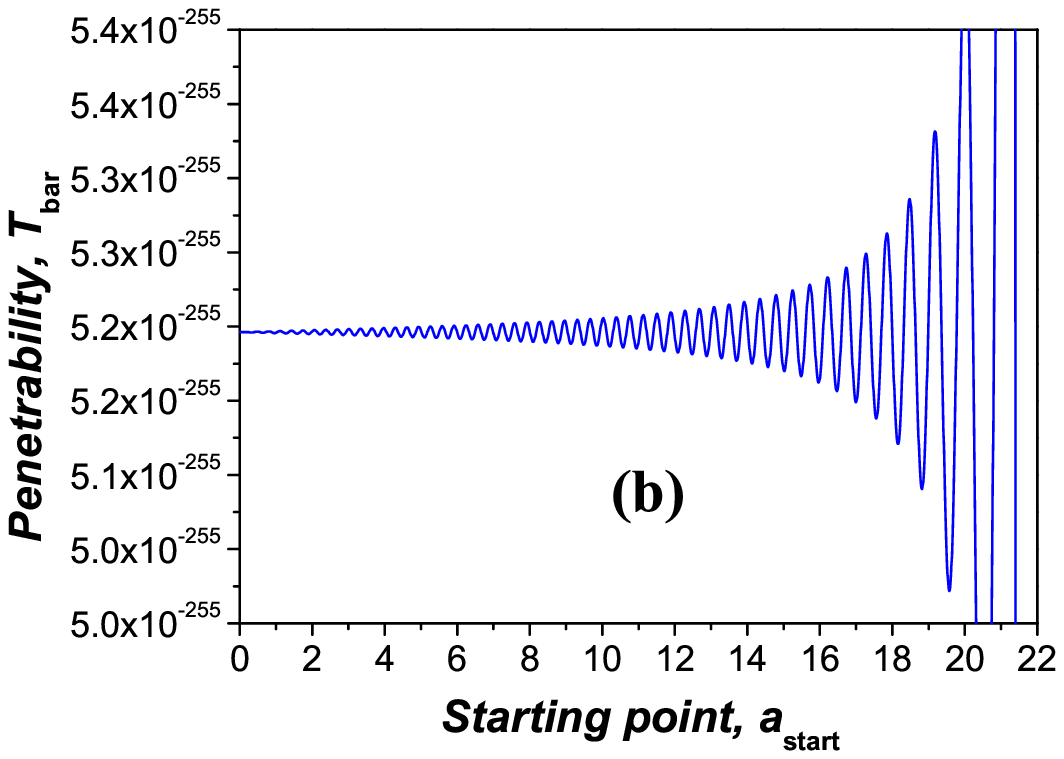}
\hspace{-7mm}\includegraphics[width=65mm]{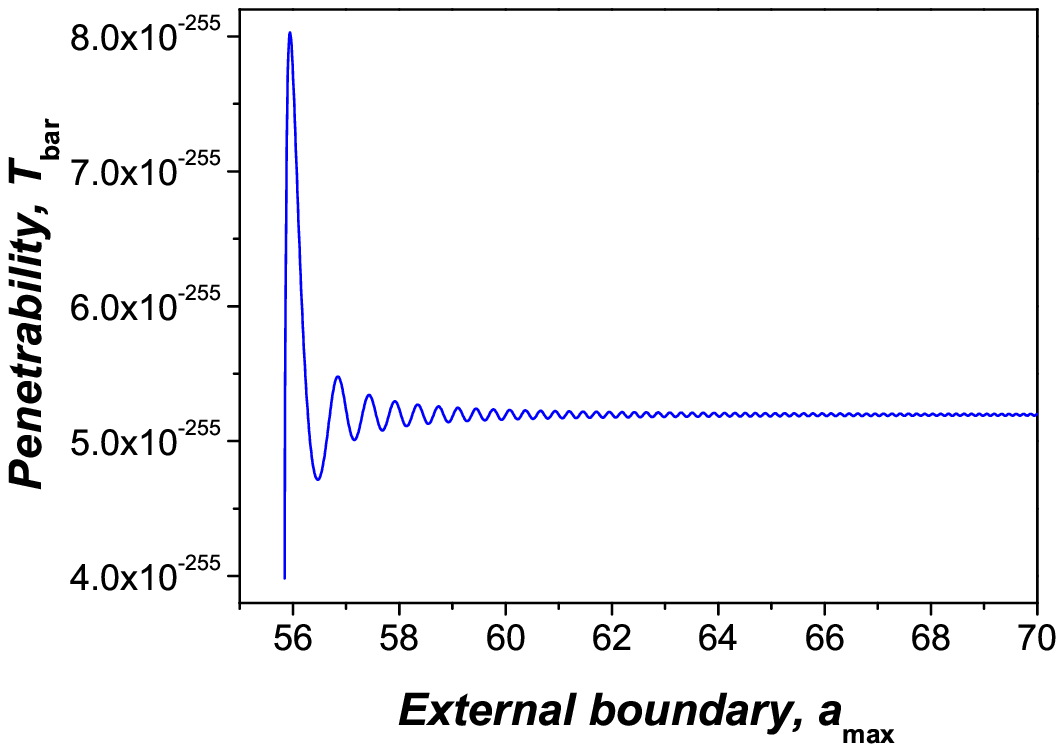}}
\vspace{-5mm}
\caption{\small (Color online)
Penetrability of the barrier for model~\cite{Maydanyuk.2011.EPJP} at $E_{\rm rad}=100$ (parameters of calculation: 10000 intervals at $a_{\rm max}=100$):
(a) dependence of penetrability on the starting point $a_{\rm start}$ in region from zero up to internal turning point, at fixed $a_{\rm max}=100$:
at increasing of $a_{\rm start}$ the penetrability oscillates, maxima are increased and minima are decreased,
(b) dependence of the penetrability on the starting point $a_{\rm start}$ in more detailed consideration,
(c) dependence of the penetrability on external boundary $a_{\rm max}$ at fixed $a_{\rm start}=0.1$: one can see that at increasing of $a_{\rm max}$ the calculated penetrability tends to some definite limit value, which we chose for further calculations and analysis.
\label{fig.2}}
\end{figure}

Before applying formalism above to the model with energy density dependent on velocity, at first we shall clarify, how this formalism is working in the model~\cite{Maydanyuk.2011.EPJP}. In the beginning, let us consider tunneling which take plase deeply under the barrier. We shall choose value for energy radiation as $E_{\rm rad}=100$.
Results of such calculations of penetrability are presented in Fig.~\ref{fig.2}.
In the first figure (a) it is shown how the penetrability is changed in dependence on the starting point $a_{\rm start}$.
One can see that it has oscillating behavior, maxima are slowly increased and minima are decreased at tending of the starting point to internal turning point (at fixed $a_{\rm max}=100$).
The second figure (b) is the same result, but in more enlarged presentation.
In the last figure (c) it is shown how the penetrability is changed at increasing of the external boundary $a_{\rm max}$, starting from the external turning point (at fixed $a_{\rm start}=0.1$, corresponding to coordinate of minimum of internal well before the barrier). From here it follows that:
(1) our approach gives convergent calculations,
(2) determination of the penetrability on the basis of shape of the barrier inside tunneling region only (like semiclassical approaches of the second order) seems to be enough far from calculated penetrability, obtained after taking into account external tail and internal shape of the well,
(3) inspite of sharp decreasing of the potential in the external region (after external turning point) at increasing of the scale factor $a$ (which decreases with evident acceleration!) calculations of penetrability are convergent and they coincide to some definite value (that allows us to speak about reliable values of the penetrability for such potentials).

\begin{figure}[htbp]
\centerline{\includegraphics[width=65mm]{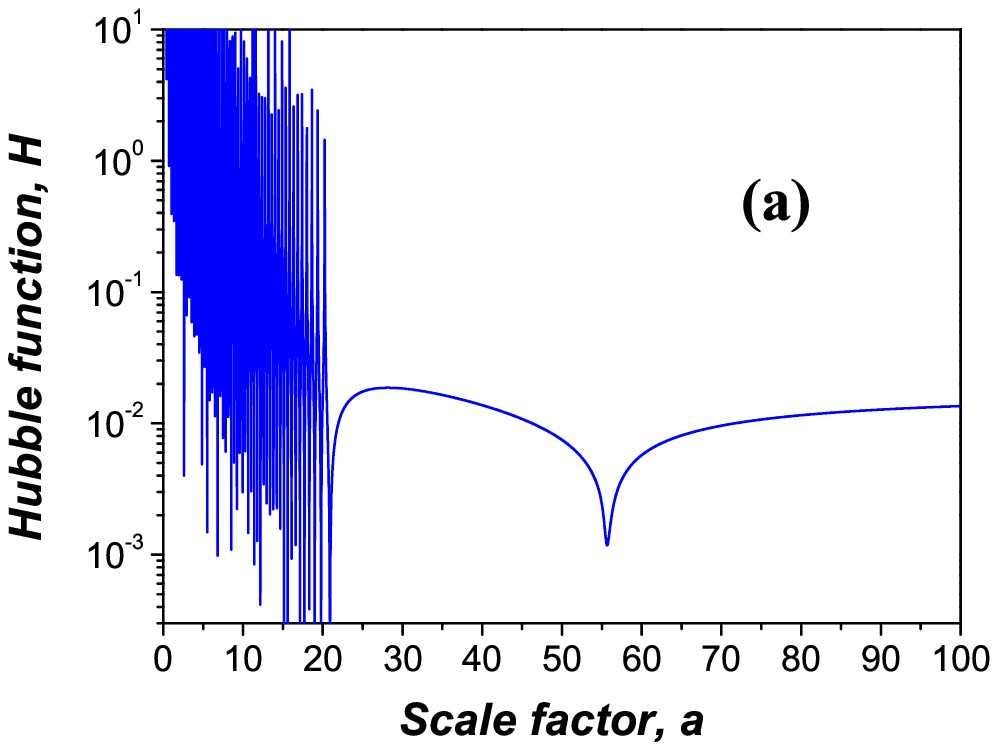}
\hspace{-7mm}\includegraphics[width=65mm]{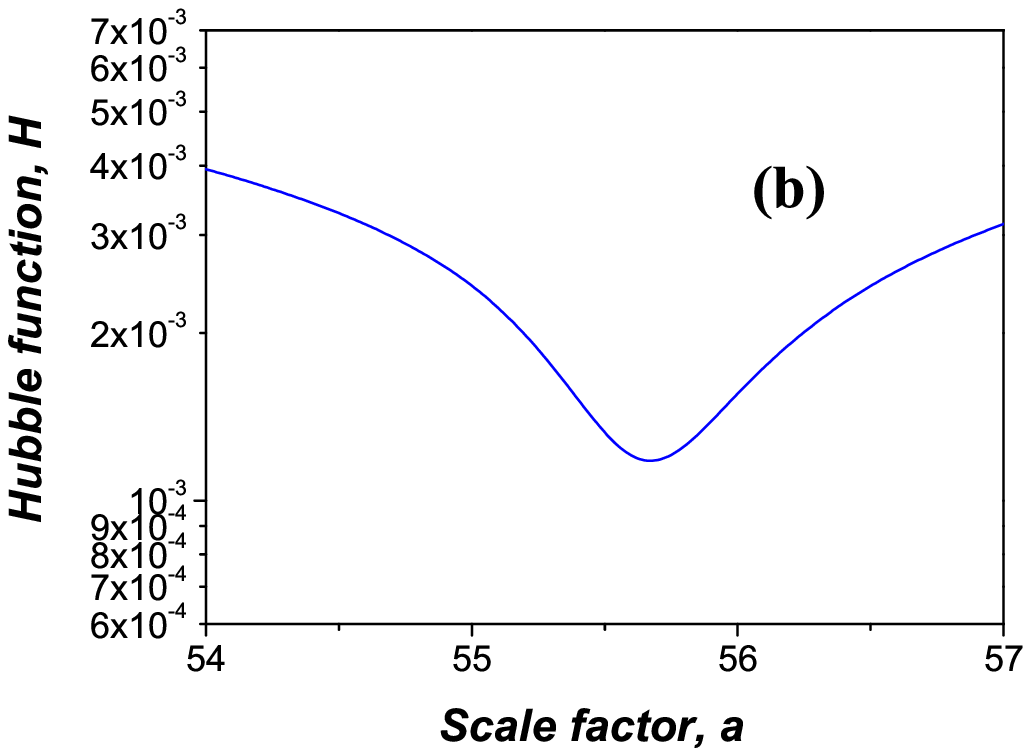}
\hspace{-7mm}\includegraphics[width=65mm]{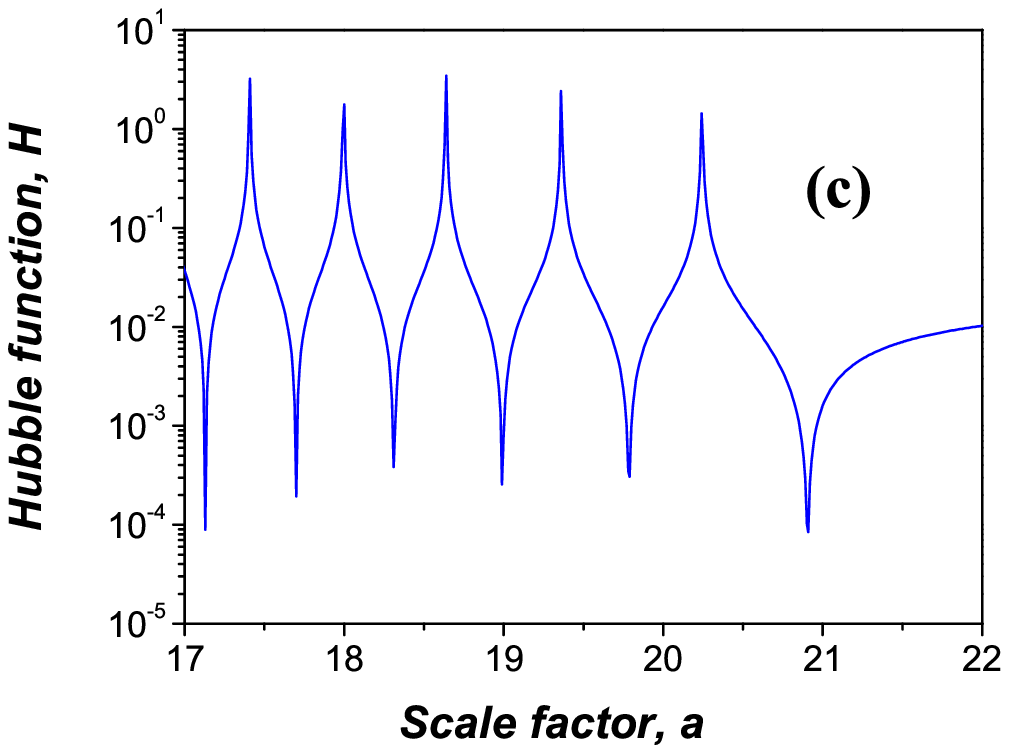}}
\vspace{-5mm}
\caption{\small (Color online)
The modulus of the function of Hubble in dependence on scale factor $a$ at $E_{\rm rad}=100$ (calculation parameters: starting point $a_{\rm start}=0.1$; 10000 intervals at $a_{\rm max}=100$):
(a) whole range of variable $a$ can be separated on 3 regions: in the internal region (at $a<21$) presence of chaotic peaks and minima is observed, in the tunneling region (at $21<a<55.6$) the function is smooth and has no any oscillation, in the external region it increases monotonously (slowly transforming to linear dependence),
(b) at transition from the tunneling region to external one smooth minimum of this function is observed, corresponding to external turning point,
(c) location of minima and peaks at small $a$ is similar (the last minimum corresponds to the internal turning point).
\label{fig.3}}
\end{figure}

Now we shall analyze the function of Hubble defined by formula~(\ref{eq.model.3.9}).
In Fig.~\ref{fig.3} it is shown how the modulus of this function is changed in dependence on scale factor $a$.
In the first figure (a) general picture is shown: at small $a$ (close to $a=21$) sharp chaotic peaks are observed, then this function has one clear maximum and minimum, at finishing it increases monotonously. In the first consideration, such a behavior of the function of Hubble (especially at small $a$) looks to be enough strange...
But after increasing insight of the last right minimum at $a=55.6$ one can see (see figure~(b)) that it is stable and is not equal to zero (that confirms coincidence and stability of computer calculations), corresponding to the external turning point --- i.e. it separates the tunneling region from the external region! In the external region the modulus of the function of Hubble increases monotonously, slowly transforming to linear dependence.
Inside the region $21<a<55.6$ the modulus of the function of Hubble has no any oscillation --- this is the tunneling region, which is finished by convergent minima at both sides.
Detailed analysis of the internal region (at $a<21$) shows that location of minima and peaks here is similar (see figure~(c)), the last minimum corresponds to the internal turning point.

\begin{figure}[htbp]
\centerline{\includegraphics[width=65mm]{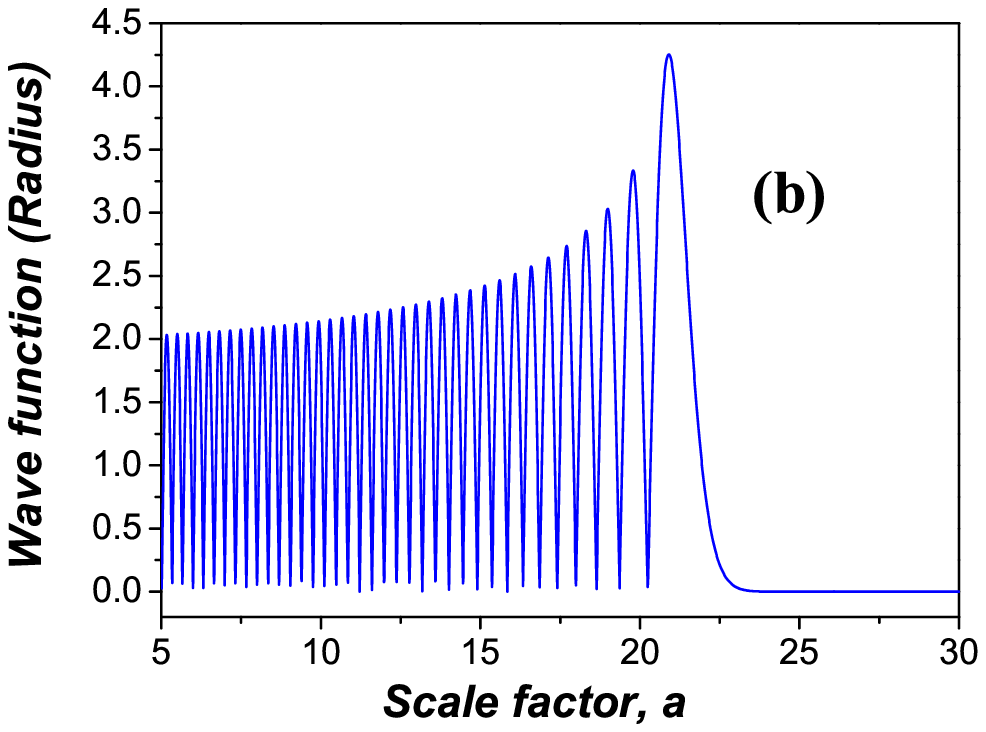}
\hspace{-7mm}\includegraphics[width=65mm]{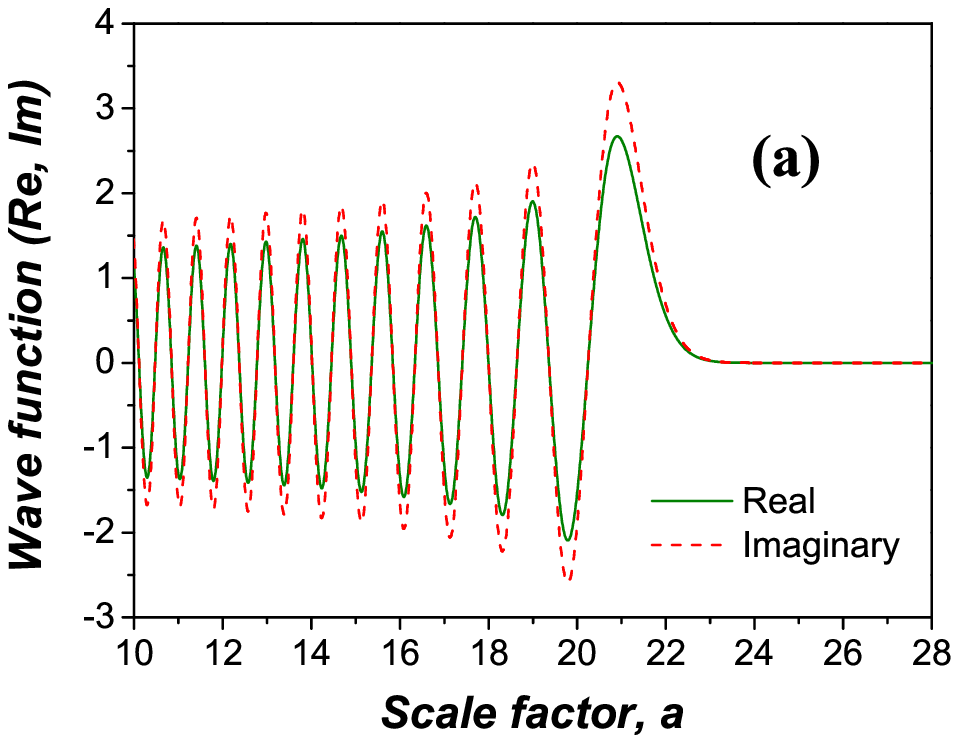}
\hspace{-7mm}\includegraphics[width=65mm]{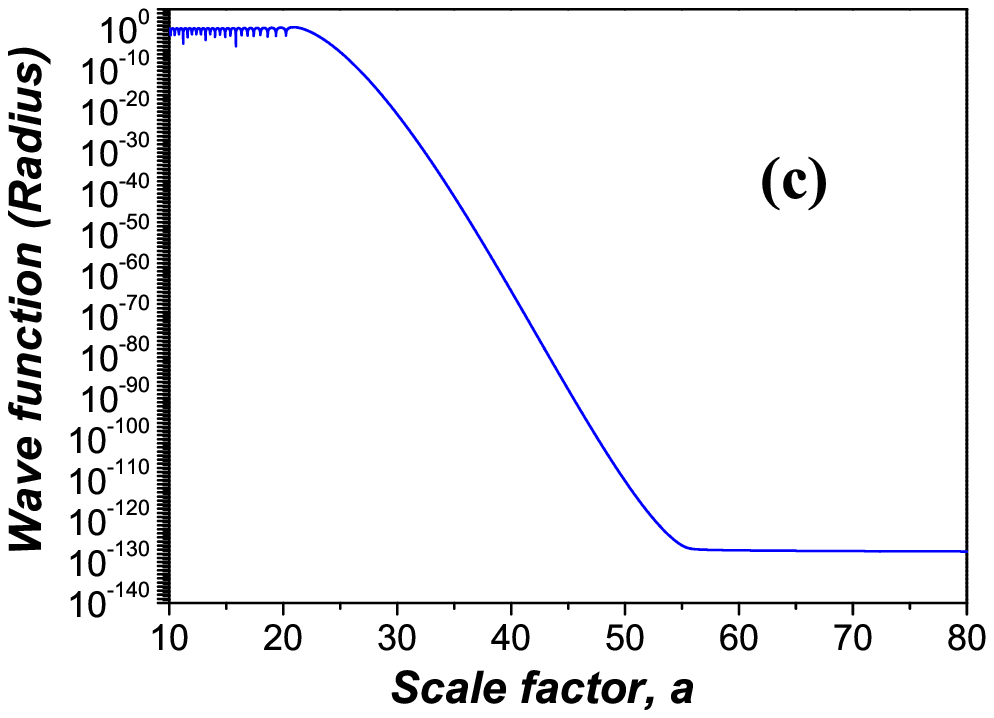}}
\vspace{-5mm}
\caption{\small (Color online)
The wave function in dependence on the scale factor $a$ at $E_{\rm rad} = 100$ (calculation parameters: the starting point $a_{\rm start} = 0.1$; 10000 intervals at $a_{\rm max} = 100$)
(a) in the internal region (at $a<21$) accelerated growth of the maxima of modulus of the wave function (with minima very closed to zero) up to the internal turning point is observed, in the tunneling and external regions (at $21<a$) this modulus decreases without oscillations,
(b) in the internal region the real and imaginary parts of the wave functions oscillate almost simultaneously, while in the tunneling and external regions they extremely fall down,
(c) consideration of the wave function in the logarithmic scale clearly demonstrates its non-zero values in the tunneling and the external regions.
\label{fig.4}}
\end{figure}

Here the following question can be appeared: whether peaks are finite at small $a$ or we are dealing with divergencies in calculations? Let us consider formula (\ref{eq.model.3.9}) for the function of Hubble: one can see that maximal values should be caused by practically zero values of the wave function, which is in the denominator.
The wave function calculated for this process is shown in Fig.~\ref{fig.4}.
In the first figure (a) the modulus of the wave function is shown: indeed, with clearly determined finite maxima (which should definitely give not zero minima of the function of Hubble) the sharp minima are seen tending to zero --- it explains presence of the sharp peaks in the Hubble function in Fig.~\ref{fig.2}. This could be explained by almost simultaneous zeroing of the real and imaginary parts of the wave function. However, this is strange as we have non-zero complex wave function (as it defines non-zero constant flux directed outside inside entire region of variable $a$), so zeros of its real and imaginary parts should not be coincide anywhere.

In the next figure (b) the real and imaginary parts of the wave function are shown. Here, one can see clear stable curves that demonstrates convergence and stability of calculations, and we need to understand the result. However, the curves behave almost similarly: their maxima, minima and zeroes are located at close coordinates. This situation is similar to the behavior of the wave function of the bound state for a particle inside a potential well (with infinitely high boundaries). Indeed, at the chosen energy the penetrability is very small ($ T_{\rm bar} \sim 10^{-255}$) and output is extremely small. By other words, we are dealing with the quasi-stationary state with extremely small output outside, which is very close to stationary one, practically.

In the last figure~(c) the modulus of the wave function is shown in the logarithmic scale. Here, one can see that it uniformly decreases inside the tunneling region. It is clear that in the external region the modulus is constant and non-zero: this confirms the convergence of the calculations, proves presence of the stable non-zero minima of the modulus of the wave function and the finite maxima of the peaks of the function of Hubble in the internal region.


\subsection{Tunneling near the barrier maximum and above-barrier propagation processes
\label{sec.results.2.2}}

Now let us consider a case where the tunneling occurs near the barrier maximum. In this case, we can use our previous analysis in~\cite{Maydanyuk.2011.EPJP} and choose $E_{\rm rad} = 220$. So, at start in point $a_{\rm start} = 0.1$ we obtain $T_{\rm bar} = 1.52129237224042 \ cdot 10^{-7}$, $R_{\rm bar} = 0.999999847870763$ and the condition $T_{\rm bar} + R_{\rm bar} = 1$ holds up to 14 digits%
\footnote{Note that the semiclassical methods usually do not give the reflection coefficient $R_{\rm bar}$ and the mentioned test is not applied. However, in discussions on comparison between fully quantum calculations and semiclassical ones this important point is usually ignored, with assumptions on advantage of the semiclassical apparatus without alternatives.}.

\begin{figure}[htbp]
\centerline{\includegraphics[width=65mm]{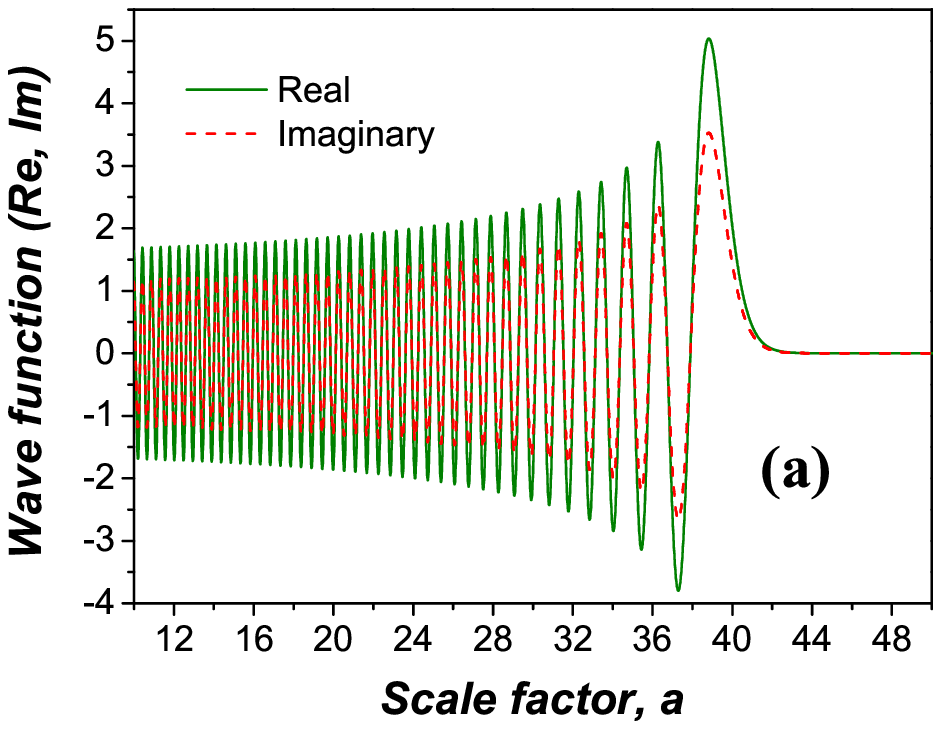}
\hspace{-7mm}\includegraphics[width=65mm]{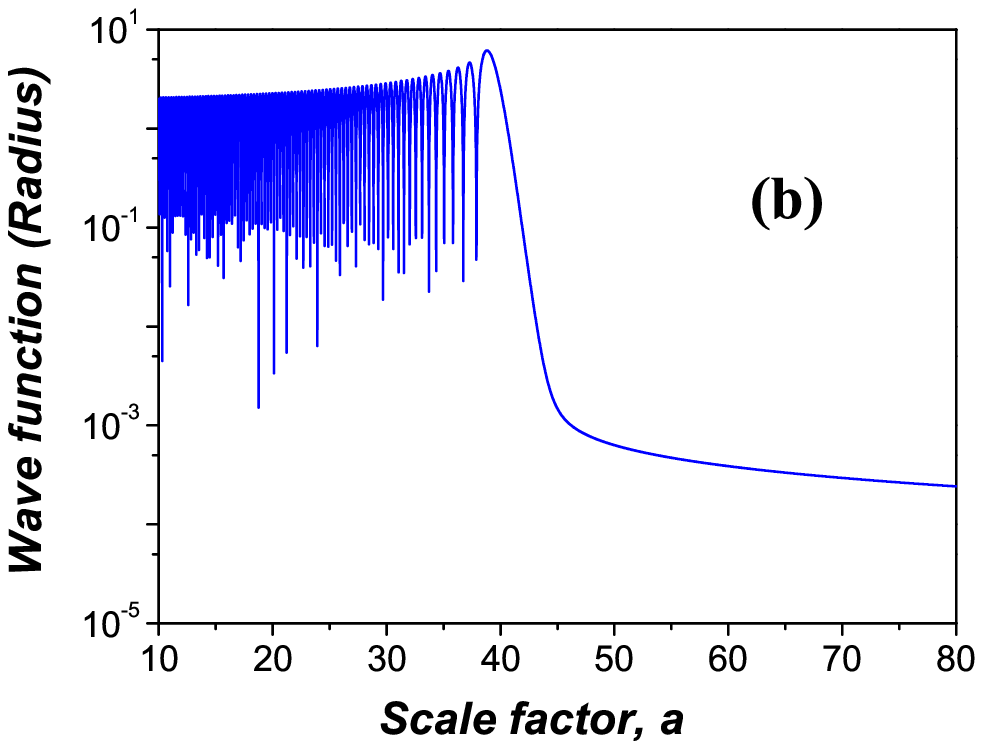}
\hspace{-7mm}\includegraphics[width=65mm]{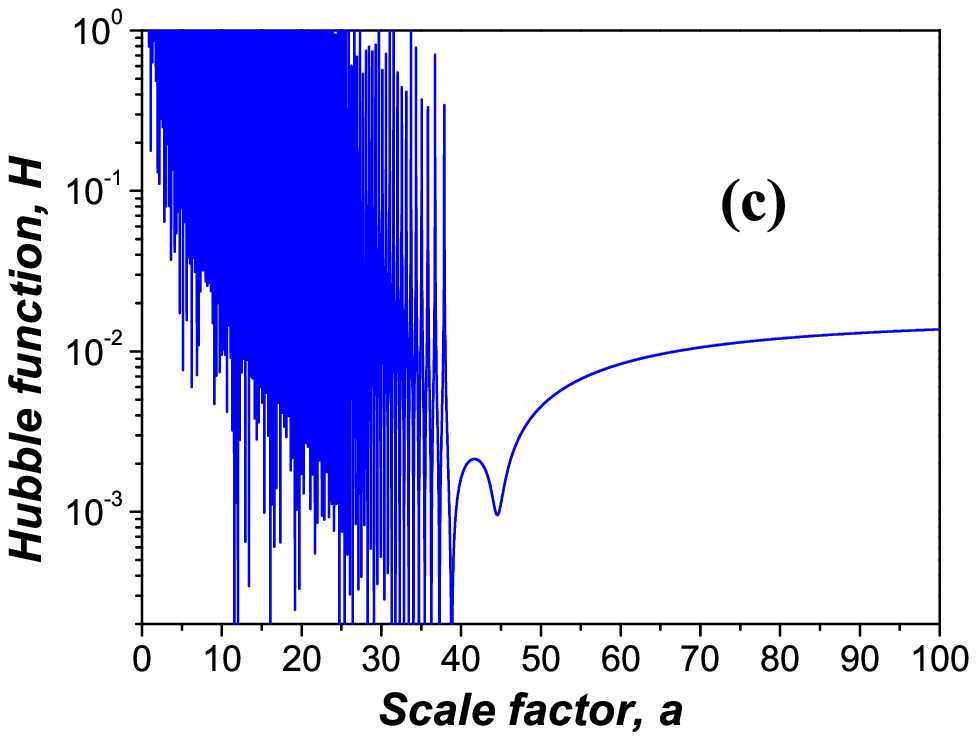}}
\vspace{-5mm}
\caption{\small (Color online)
The wave function and the function of Hubble in dependence on the scale factor $a$ at $E_{\rm rad} = 220$ (calculation parameters: the starting point $a_{\rm start} = 0.1$, 10000 intervals at $a_{\rm max} = 100$):
(a) in the internal region number of oscillations of the wave function is much larger and its maxima increase stronger in comparison with results at $E_{\rm rad} = 100$ (see Fig.~\ref{fig.4}~(b)), that is explained by increasing of the energy $E_{\rm rad}$ and enlarging of the internal well region;
(b) in the tunneling and external regions the modulus of the wave function is changed essentially weaker in comparison with results at $E_{\rm rad} = 100$ (see Fig.~\ref{fig.4}~(c)), that indicates on essential oncoming of energy $E_{\rm rad}$ to the barrier maximum;
(c) the function of Hubble behaves similarly to the case at $E_{\rm rad} = 100$ (see Fig.~\ref{fig.3}), but here the tunneling region is much smaller and the function of Hubble in it is smaller.
\label{fig.5}}
\end{figure}

Our calculations show that the penetrability in dependence on the starting point $a_{\rm start}$ and on the external boundary $a_{\rm max}$ behaves like to the case studied above at $E_{\rm rad} = 100$ (also see~\cite{Maydanyuk.2011.EPJP}).
Placing the starting point in the minimum of the internal well, we calculate the wave function and the function of Hubble (see Fig.~\ref{fig.5}).
From these figures it is clear that these functions behave as obtained above at $E_{\rm rad} = 100$. However, in this case we observe:
(1) in the internal region number of oscillations of the wave function is essentially larger and the difference between its maxima is reinforced (see Fig.~\ref{fig.5}~(a) in comparison with Fig.~\ref{fig.4}~(b)), caused by enlarging of the internal region with shift of the internal turning point to the right;
(2) in the tunneling and external regions the difference between the maxima and minima of modulus of the wave function is much smaller (see Fig.~\ref{fig.5}~(b) in comparison with Fig.~\ref{fig.4}~(c)), that indicates on essential oncoming of the energy $E_{\rm rad}$ to the barrier maximum and much stronger outgoing flux through the barrier outside;
(3) the tunneling region is less, which can be easily found by typical behavior of the function of Hubble (see Fig.~\ref{fig.5}~(c)).

\begin{figure}[htbp]
\centerline{\includegraphics[width=65mm]{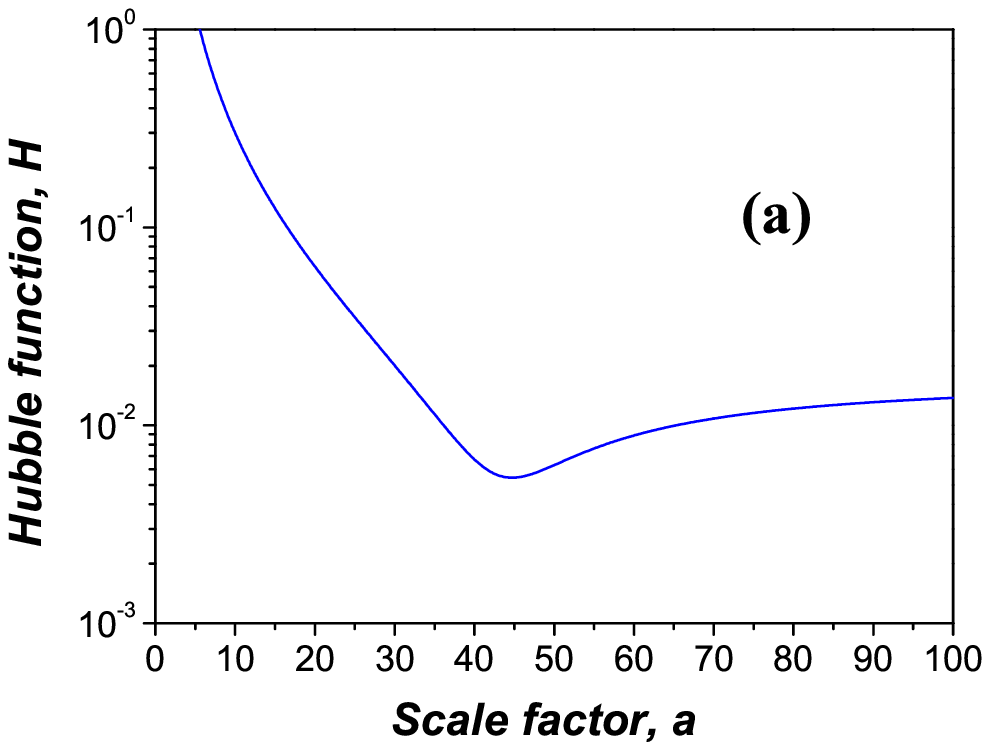}
\hspace{-7mm}\includegraphics[width=65mm]{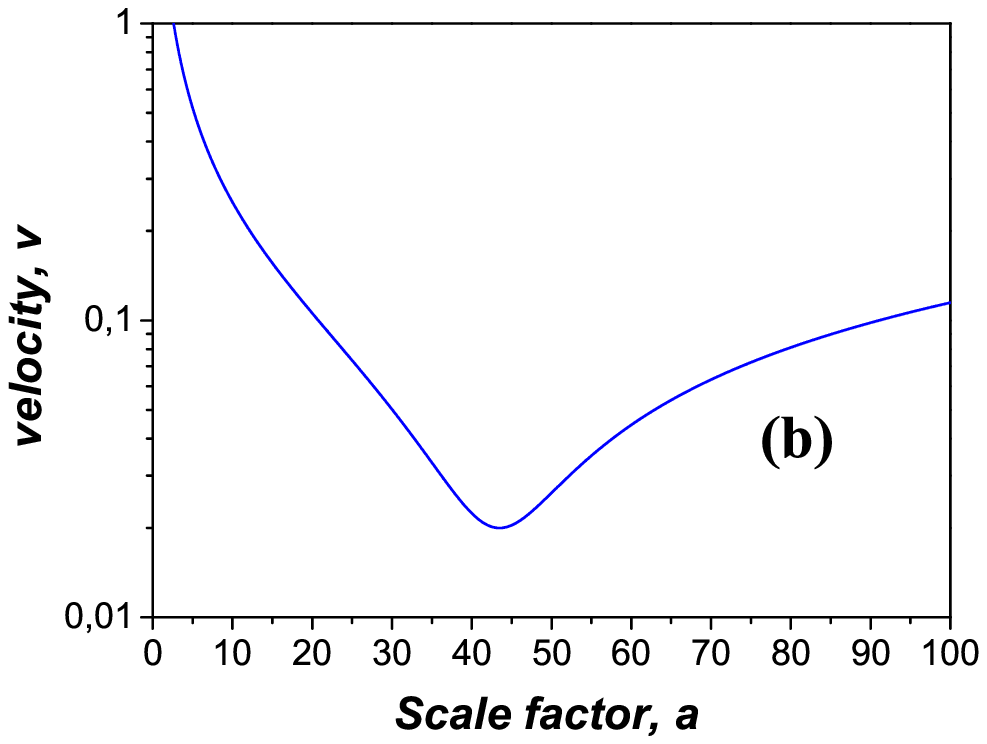}
\hspace{-7mm}\includegraphics[width=65mm]{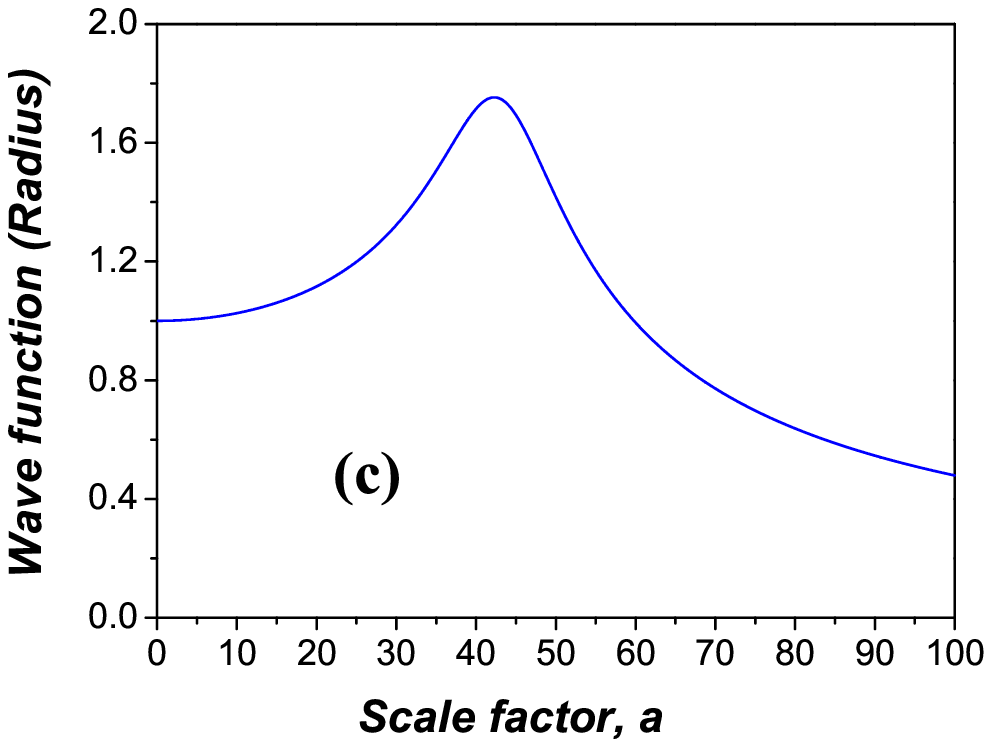}}
\vspace{-5mm}
\caption{\small (Color online)
The function of Hubble and wave function in dependence on the scale factor $a$ at $E_{\rm rad} = 250$ (calculation parameters: the starting point $a_{\rm start} = 0.1$, 10000 intervals at $a_{\rm max} = 100$):
(a) the function of Hubble has a smooth form: in the internal region the sharp peaks and minima (typical for above-barrier energies and reflecting bound state) are disappeared completely, in the middle region there is only one smooth minimum in the coordinate of the barrier maximum (tunneling maximum is disappeared as shown in Figs.~\ref{fig.3}~(a) and \ref{fig.5}~(c)), in the external region the accelerated increase of the function of Hubble remains (as in Figs.~\ref{fig.3}~(a) and \ref{fig.5}~(c));
(b) the wave function is oscillatory in the entire region;
(c) modulus of the wave function has maximum in coordinate of the barrier maximum.
\label{fig.6}}
\end{figure}

Now let us find out how these pictures will be changed, if we increase the energy $E_{\rm rad}$ above the barrier.
Results such calculations are shown in Fig.~\ref{fig.6}, where we have chosen $E_{\rm rad} = 250$.
In the first figure (a) the function of Hubble is shown. One can see that
(1) In the internal region the sharp peaks and minima have disappeared completely,
(2) In the middle region the behavior of this function, typical for the tunneling region, has disappeared also, and instead there is a smooth minimum in the coordinate of the potential barrier maximum,
(3) In the external region the monotonic increasing of the function of Hubble remains, slowly becoming to linear dependence.
The wave function and its modulus are shown in next two figures~(b,~c). In this case, we see oscillatory behavior, typical for the above-barrier energies. However, its maximum is located at coordinate of the barrier maximum, with a monotonic decrease to both sides.

\begin{figure}[htbp]
\centerline{\includegraphics[width=65mm]{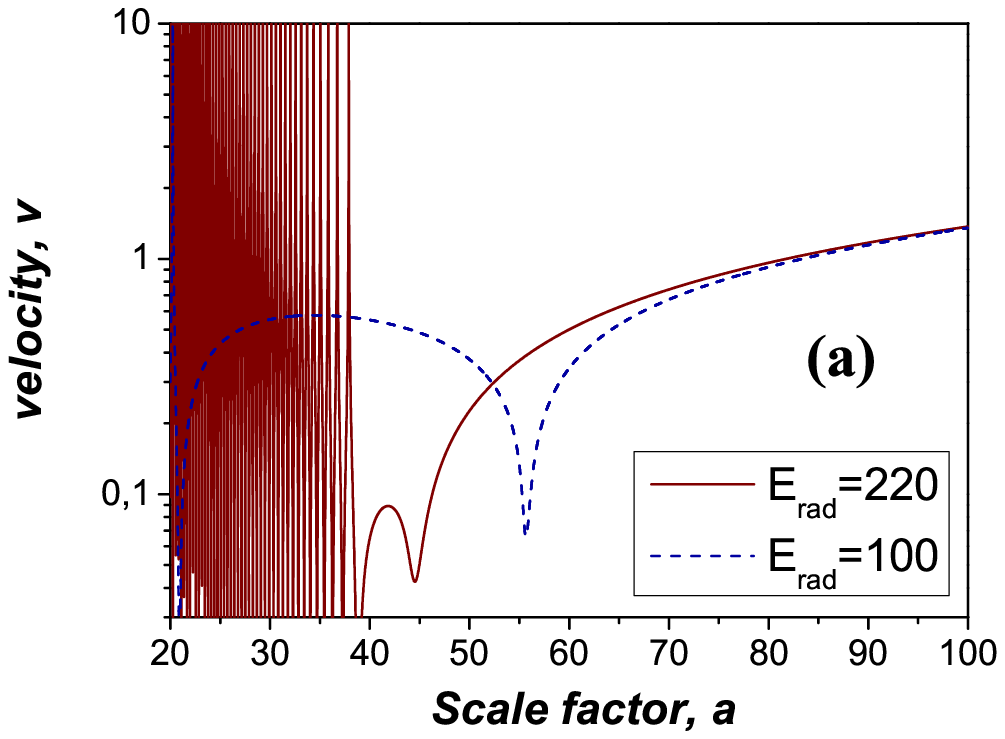}
\hspace{-7mm}\includegraphics[width=65mm]{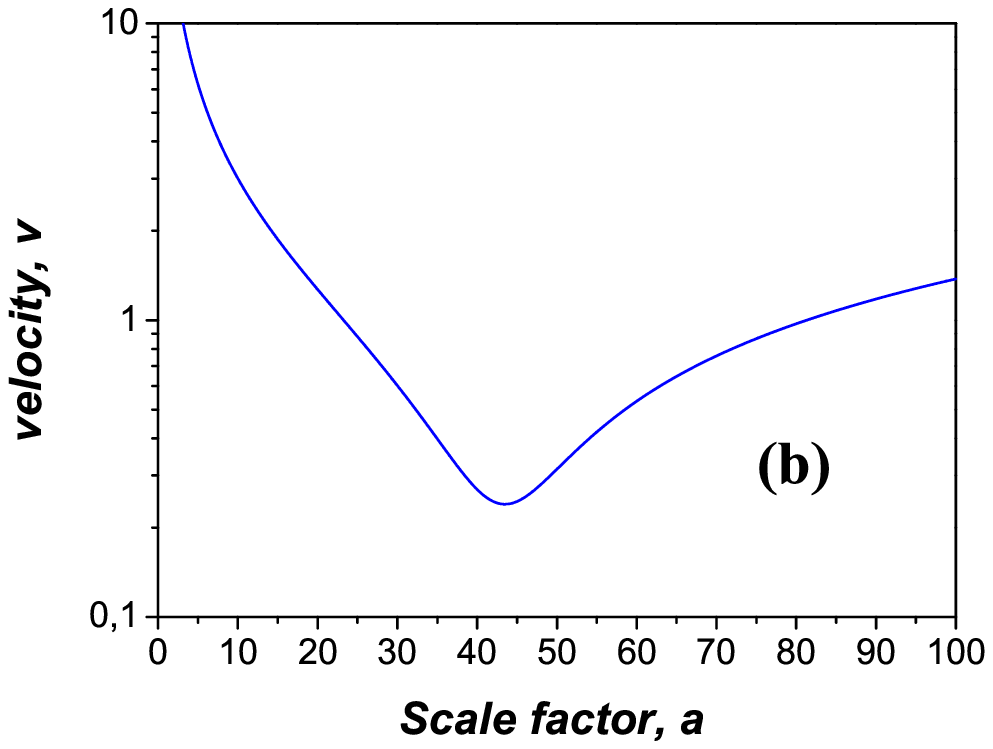}
\hspace{-7mm}\includegraphics[width=65mm]{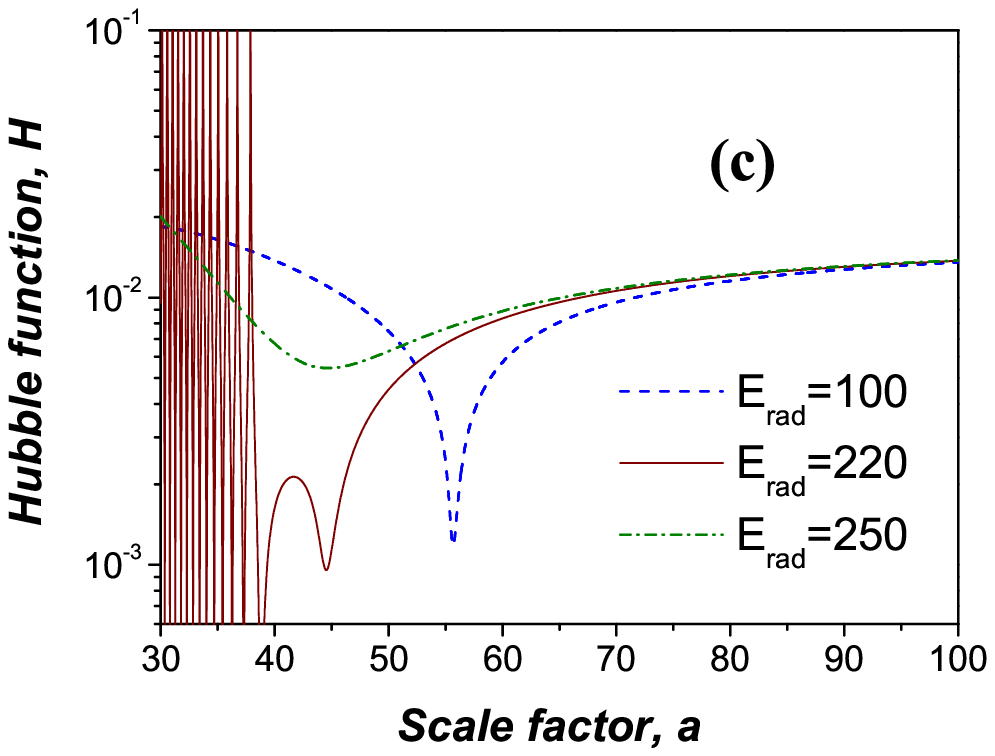}}
\vspace{-5mm}
\caption{\small (Color online)
The velocity of expansion of universe in dependence on the scale factor $a$ (calculation parameters: the starting point $a_{\rm start} = 0.1$, 10000 intervals at $a_{\rm max} = 100$):
(a) in the internal region at sub-barrier energies there are sharp peaks and minima, in the tunneling region the velocity is smooth, in the external region velocities at different energies tend to the same limit,
(b) for the above-barrier energy $E_{\rm rad} = 250$ sharp peaks and minima are disappeared, hump of tunneling (previously observed for the sub-barrier energies) transforms to smooth minimum in coordinate, corresponding to the potential barrier maximum,
(c) the functions of Hubble at different energies $E_{\rm rad}$ tend to the same limit at large $a$.
\label{fig.7}}
\end{figure}

Now let us analyze, how velocity defined by the formula (\ref{eq.model.3.9}) behaves in each case.
Such calculations are presented in Fig.~\ref{fig.7}.
In general, the behavior of the velocity looks like the function of Hubble. From the figures it is clear that if in the internal region there are sharp peaks and minima at the sub-barrier energies (see figure~(a)), then they disappear at increasing of the energy $E_{\rm rad}$ above the barrier (see figure~(b)). However, after detailed analysis one can see the oscillations with a general tendency to decreasing (and decreasing amplitude with increasing $E_{\rm rad}$). They can be explained by wave nature, which quantum-mechanical treatment of the process gives us. They are the previously considered sharp peaks and minima at sub-barrier energies, i.e. now picture of sub-barrier and above-barrier processes becomes unite. Figure~(c) shows that the function of Hubble at different energies $E_{\rm rad}$ tend to the same limit behavior at large $a$.

So, there is the following picture at the above-barrier energy $E_{\rm rad} = 250$. We have no any classically forbidden region inside whole area of the scale factor $a$. Hence it would seem that this situation forbids to consider formation of the universe, and we would reject this case. However, we see that for values of $a$ less than coordinate of the barrier maximum (we denote it as $a_{\rm bar}$) the velocity gradually decreases to minimum with increasing of $a$. In this case, the modulus of the wave function increases up to maximum: it points to a gradual increase of probability of appearance of the universe, with maximum at $a_{\rm bar}$. If to take the spherically symmetric picture of the extension into account (where the density of matter in filled volume should not increase and it can be associated with the probability of appearance of the the universe), then such an increase of the probability is contrary to natural expansion of the universe in classical treatment. So, this situation is more suitable to description of formation of the universe with classical space-time, with its birth at $ a_{\rm bar}$. Starting from $a_{\rm bar}$, the expansion of space-time becomes classic. Before to this coordinate, the universe is formed with a gradual damping expansion, at maximum in point $a_{\rm bar}$ this extension is practically stops for a certain period, then further expansion begins with the acceleration. This logic points us to competence of quantum description of formation of the universe at the above-barrier energies (when there is no tunneling).

Also another property is found:
\emph{All velocities at different energies $E_{\rm rad}$ tend to the same limit with increasing $a$ in the external region: i.e. this model gives the same dynamics of the accelerated expansion of the universe for later times with completely different scenarios of its evolution in the first stage.}

\section{Model with energy density dependent on velocity of expansion
\label{sec.results.3}}

\begin{figure}[htbp]
\centerline{
\includegraphics[width=85mm]{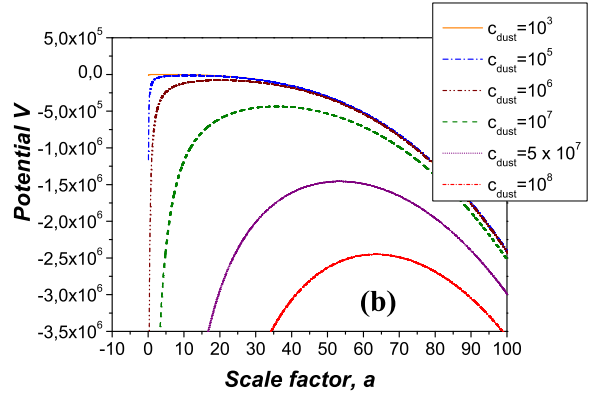}}
\vspace{-2mm}
\caption{\small (Color online)
Inclusion of the density component dependent on the velocity into the model allows to form the barrier at $k = 0$:
One can see that with increase of parameter $c_{\rm dust}$ the barrier is shifted to the right down, forming more rounded shape (calculation parameters: $E_{\rm rad} = 100$,
solid orange line is for $c_{\rm dust} = 10^{3}$,
dash-dotted blue line for $c_{\rm dust} = 10^{5}$
dash-double dotted brown line for $c_{\rm dust} = 10^{6}$,
dashed green line for $c_{\rm dust} = 10^ {7} $
short dashed purple line for $c_{\rm dust} = 5 \cdot 10^{7}$
short dash-dotted red line for $c_{\rm dust} = 10^{8}$).
\label{fig.1}}
\end{figure}

Now we shall analyze the behavior of the model with energy density dependent on the velocity at $k = 0$ and $n = 2$. The potentials generated in this approach are shown in Fig.~\ref{fig.1}: one can see that the energy density depending on the velocity forms the barrier for different values of the potential parameters.
In next Figs.~\ref{fig.8} the results of our calculations of the function of Hubble, velocity and the modulus of the wave function are presented.
\begin{figure}[htbp]
\centerline{\includegraphics[width=65mm]{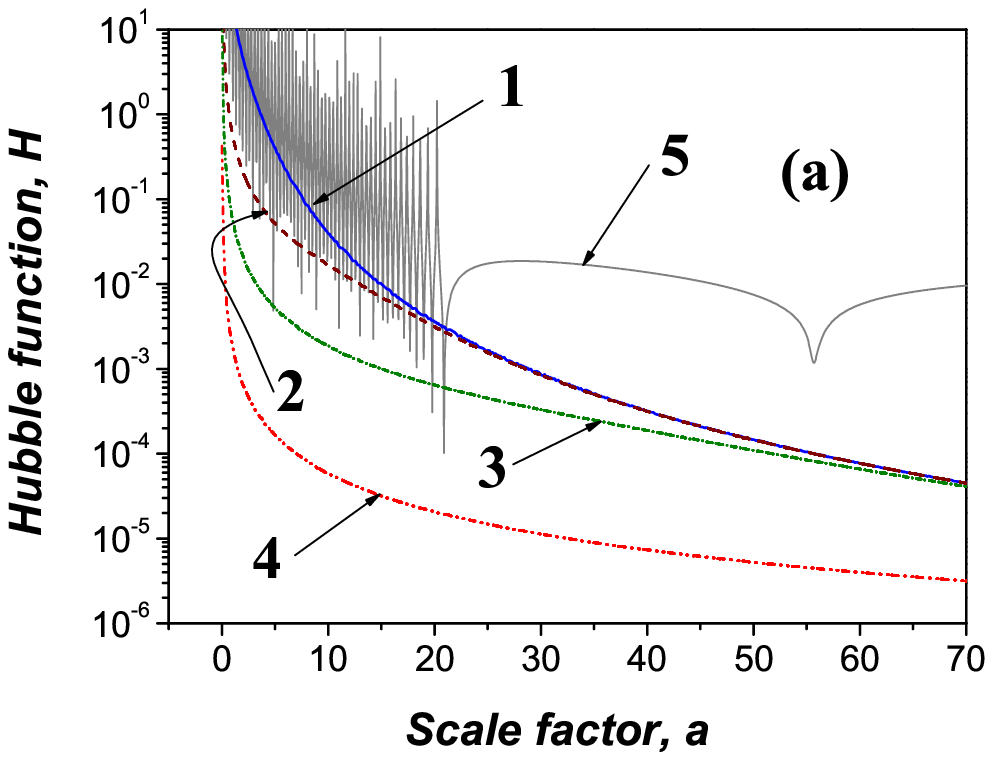}
\hspace{-7mm}\includegraphics[width=65mm]{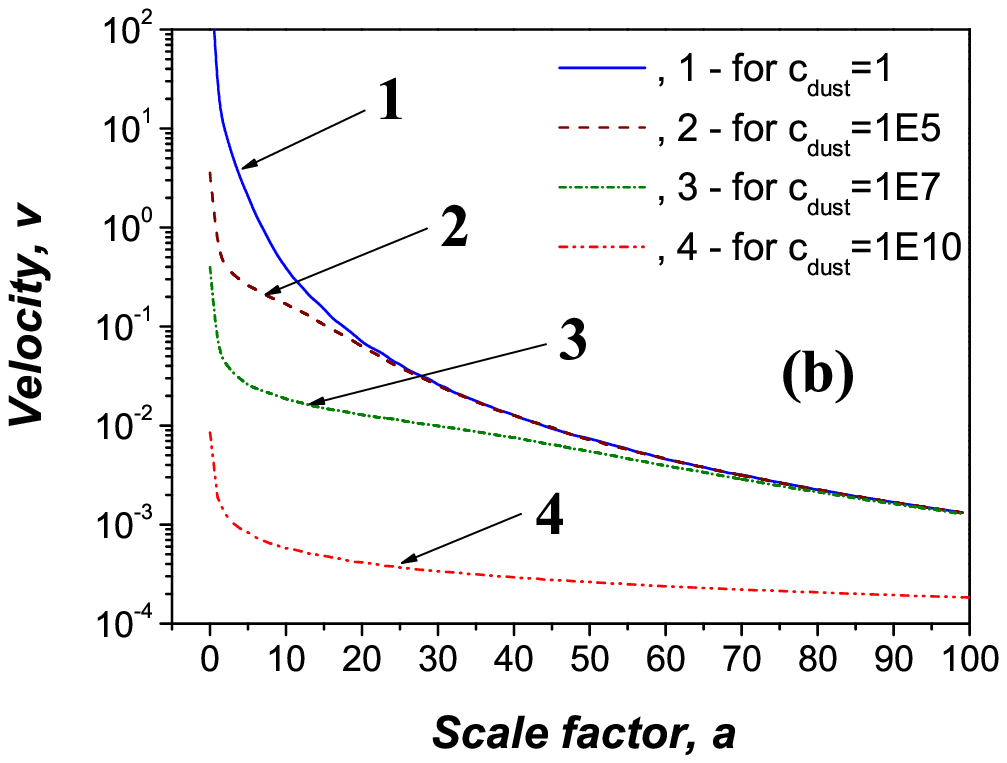}
\hspace{-7mm}\includegraphics[width=65mm]{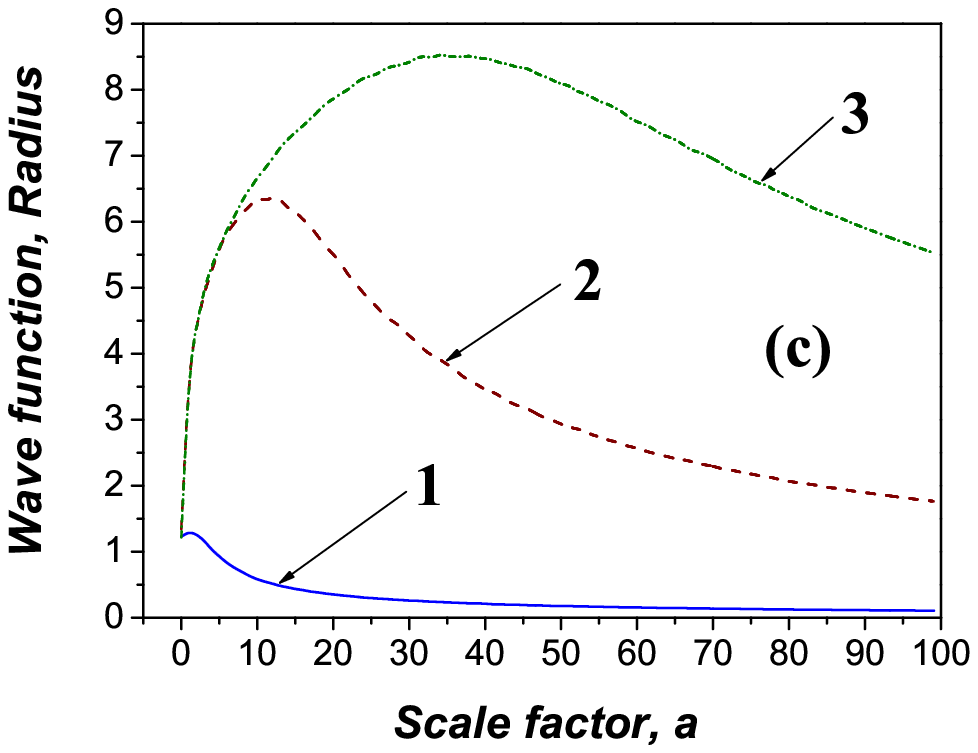}}
\vspace{-5mm}
\caption{\small (Color online)
The function of Hubble, velocity and the wave function in dependence on the scale factor $a$ at $E_{\rm rad} = 100$ and $k = 0$ for the model with velocity-dependent density
(calculation parameters: the starting point $a_{\rm start} = 0.01$, 10000 intervals at $a_{\rm max} = 100$;
blue solid line 1 is for $c_{\rm dust}=1$,
brown dashed line 2 for $c_{\rm dust}=10^{5}$,
green shot dash-dotted line 3 for $c_{\rm dust}=10^{7}$,
red dash-double dotted line 4 for $c_{\rm dust}=10^{10}$):
(a) the function of Hubble decreases monotonically at increase of the scale factor $a$ and parameter $c_{\rm dust}$
(curve 5 is added on this figure for the model~\cite {Maydanyuk.2011.EPJP}, which is much higher and tends to higher limit);
(b) for different values of $c_{\rm dust}$ the velocity tends to the same limit (at $c_{\rm dust} = 10^{10}$ this coincidence occurs at much higher $a$),
(c) the modulus of the wave function is similar to the barrier shape, maximum of this modulus corresponds to the coordinate of the barrier maximum.
\label{fig.8}}
\end{figure}
One can see that the universe is slower expanding in frameworks of such a model, with increase of the scale factor $a$ its velocity decreases, tending to some limit monotonic dependence. So, the model with the density dependent on the velocity gives slow expansion at late times, in contrast to the model~\cite{Maydanyuk.2011.EPJP}.

\begin{figure}[htbp]
\centerline{\includegraphics[width=65mm]{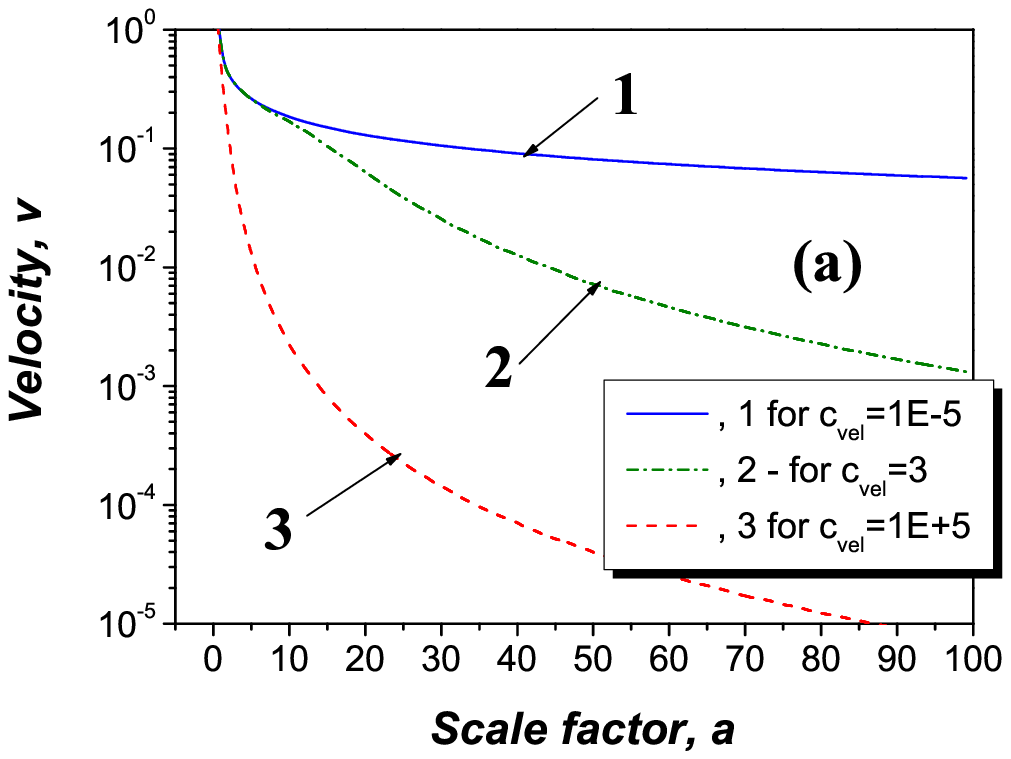}
\hspace{-7mm}\includegraphics[width=65mm]{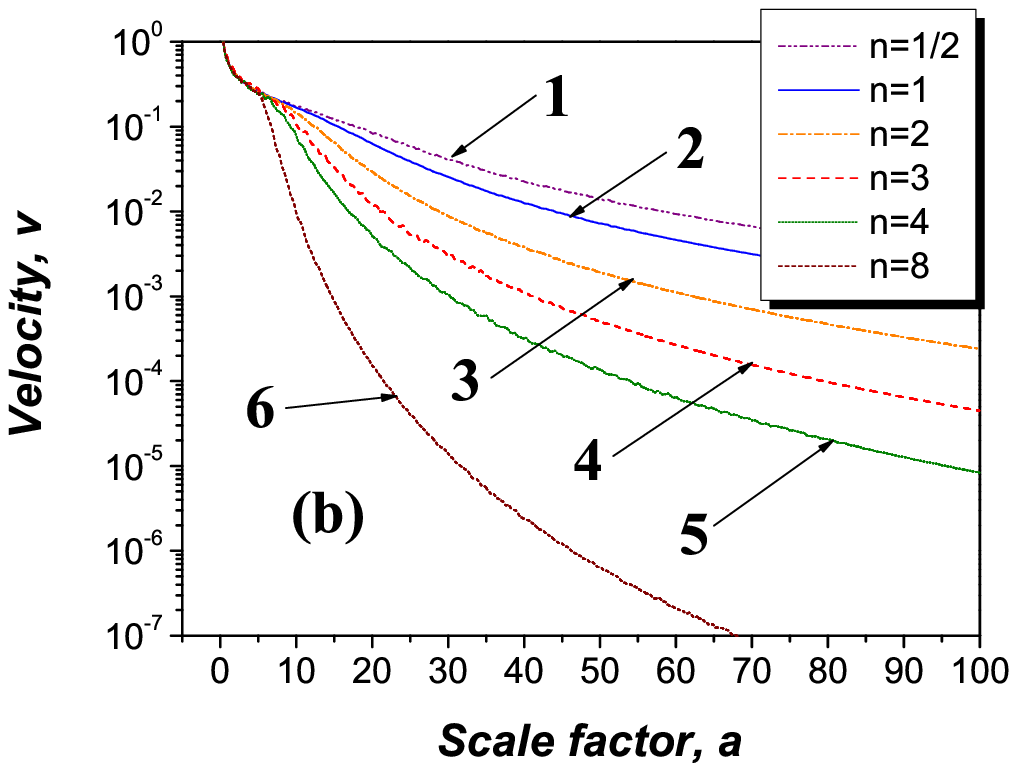}
\hspace{-7mm}\includegraphics[width=65mm]{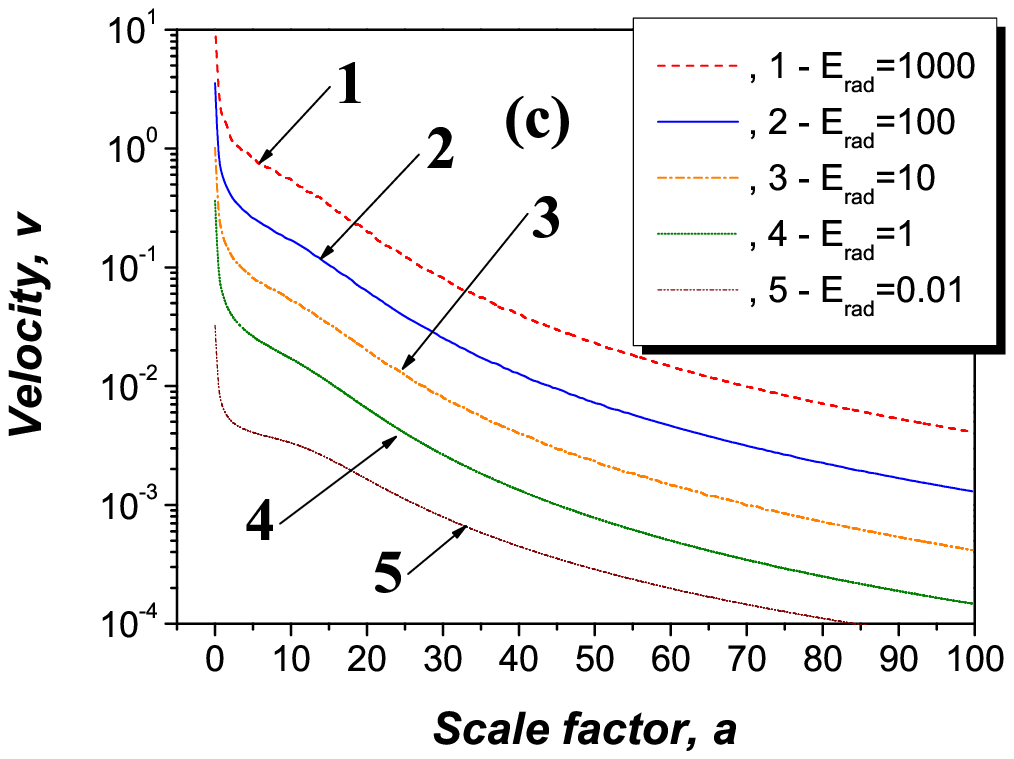}}
\vspace{-5mm}
\caption{\small (Color online)
The velocity of expansion of the universe in dependence on the scale factor $a$ for the model with component of the energy density dependent on the velocity at $k = 0$ and variations of parameters $c_{\rm vel}$, $n$ and $E_{\rm rad}$
(calculation parameters: the starting point $a_{\rm start} = 0.01$, 10000 intervals at $a_{\rm max} = 100$):
(a) calculations for different $c_{\rm vel}$ at fixed $c_{\rm dust} = 10^{5}$, $E_{\rm rad} = 100$ and $n = 1$
(blue solid line 1 is for $c_{\rm vel}=10^{-5}$,
brown dash-dotted line 2 for $c_{\rm vel}=1$,
green dashed line 3 for $c_{\rm vel}=10^{5}$);
(b) calculations for different powers $n$ at fixed $c_{\rm dust} = 10^{5}$, $c_{\rm vel} = 3$ and $E_{\rm rad} = 100$
(purple dash-double dotted line 1 is for $n=1/2$,
blue solid line 2 for $n=1$,
orange dash-dotted line 3 for $n=2$,
red dashed line 4 for $n=3$,
green short dotted line 5 for $n=4$,
wine short-dashed line 6 for $n=8$);
(c) calculations for different energies $E_{\rm rad} = 100$ at fixed $c_{\rm dust} = 10^{5}$, $c_{\rm vel} = 3$ and $n = 1$
(red dashed line 1 is for $E_{\rm rad}=1000$,
blue solid line 2 for $E_{\rm rad}=100$,
orange dash-dotted line 3 for $E_{\rm rad}=10$,
green short-dotted line 4 for $E_{\rm rad}=1$,
wine dash-double dotted line 5 for $E_{\rm rad}=0.01$).
\label{fig.9}}
\end{figure}

In Fig.~\ref{fig.9} the velocity of expansion of the universe for this model in dependence on the scale factor $a$ at variations of parameters $c_{\rm vel}$, $n$ and $E_{\rm rad}$ is shown.
In next Figs.~\ref{fig.10} we demonstrate comparing calculations for the velocity of evolution of the expansion of the universe by this model with the included cosmological constant (at $k = 0$).
Our results of calculation of the duration of the universe defined by (\ref{eq.model.4.1}) with different parameters are presented in last Fig.~\ref{fig.11}.
%
%
\begin{figure}[htbp]
\centerline{\includegraphics[width=65mm]{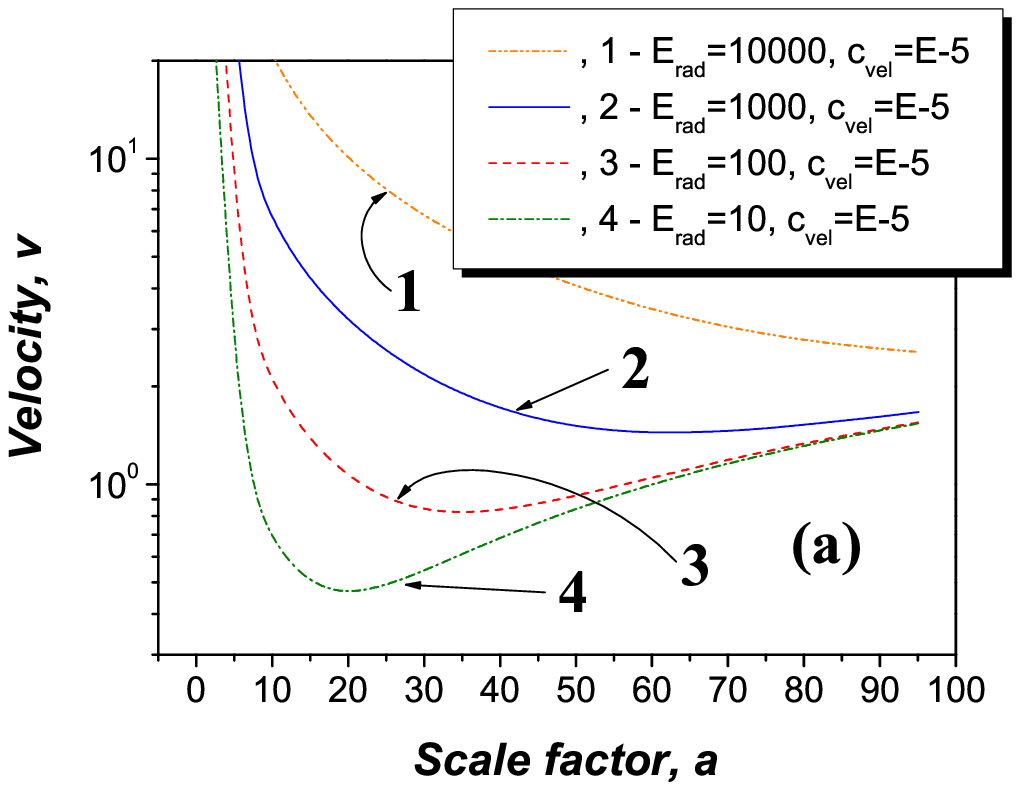}
\hspace{-1mm}\includegraphics[width=70mm]{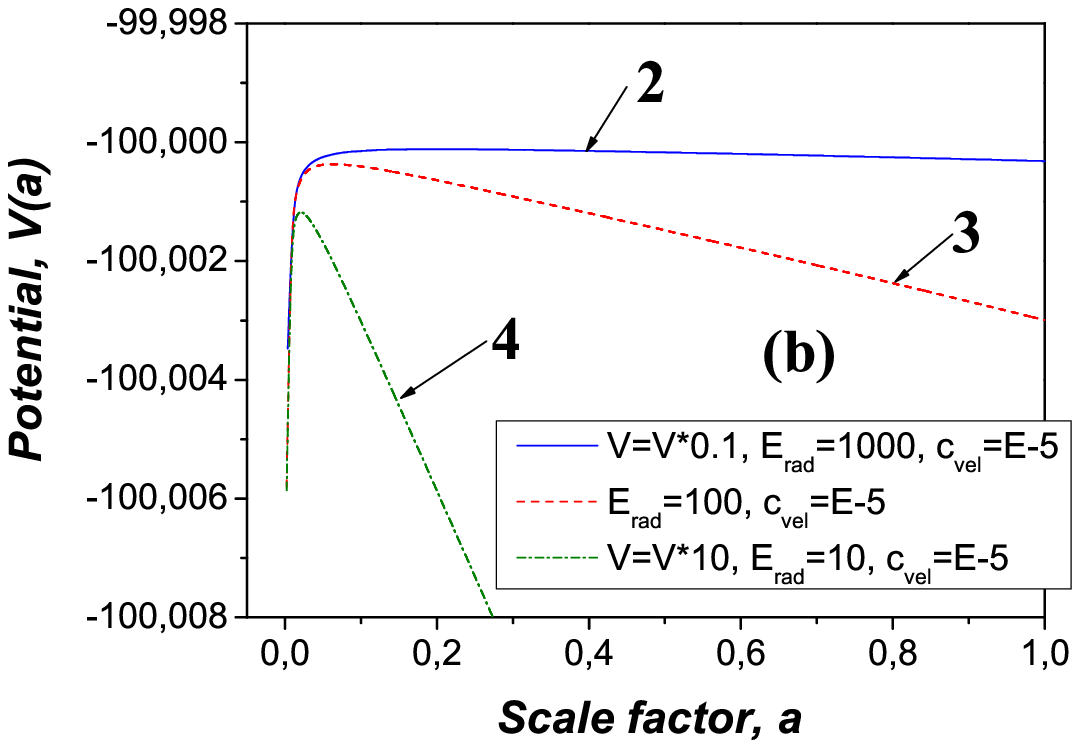}}
\vspace{-5mm}
\caption{\small (Color online)
The evolution of the expansion of the universe for the model with the component of energy density dependent on the velocity with the included cosmological constant at $k = 0$, $c_{\rm dust} = 10^{-5}$ and $c_{\rm vel} = 10^{-5}$
(calculation parameters: the starting point $a_{\rm start} = 0.01$, 10000 intervals at $a_{\rm max} = 100$;
orange dash-double dotted line 1 is for $E_{\rm rad}=10000$,
blue solid line 2 for $E_{\rm rad}=1000$,
red dashed line 3 for $E_{\rm rad}=100$,
green dash-dotted line 4 for $E_{\rm rad}=10$):
(a) velocity: varying the energy $E_{\rm rad}$, one can obtain a smooth transition from the accelerated expansion to decelerated one for later times;
(b) the potential: one can see a presence of the barrier for different parameters.
\label{fig.10}}
\end{figure}
%
%
\begin{figure}[htbp]
\centerline{\includegraphics[width=65mm]{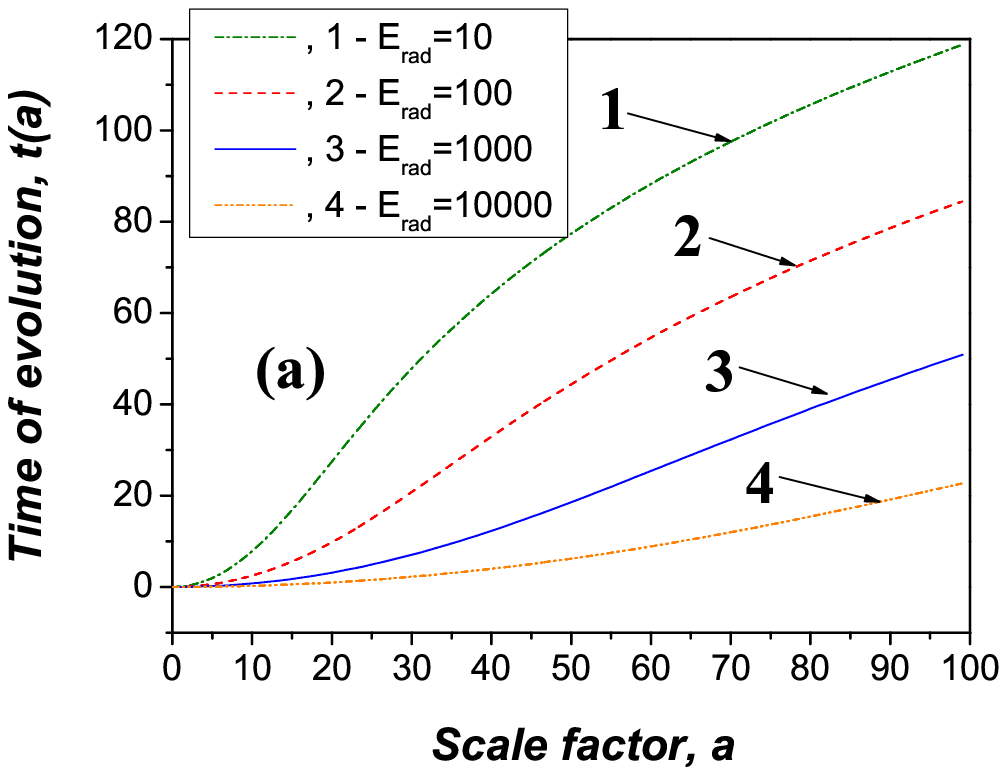}
\hspace{-1mm}\includegraphics[width=65mm]{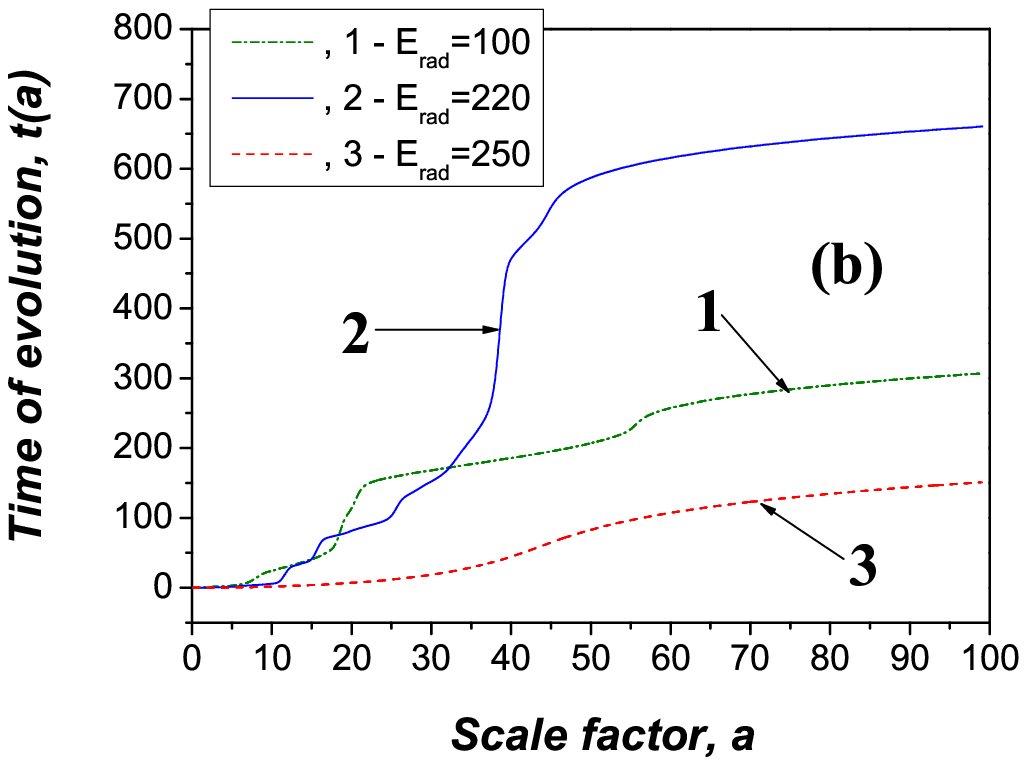}}
\vspace{-5mm}
\caption{\small (Color online)
Duration of evolution of the universe (its age) as function of the scale factor
(calculation parameters: the starting point $a_{\rm start} = 0.01$, 10000 intervals at $a_{\rm max} = 100$).
(a) Model with component of energy density dependent on the velocity with the included cosmological constant at $k = 0$, $c_{\rm dust} = 10^{-5}$ and $c_{\rm vel} = 10^{-5}$
(green dash-dotted line 1 is for $E_{\rm rad}=10$,
red dashed line 2 for $E_{\rm rad}=100$,
blue solid line 3 for $E_{\rm rad}=1000$,
orange dash-double dotted line 4 for $E_{\rm rad}=10000$):
we observe increase of the duration at decrease of the energy $E_{\rm rad}$ (for the above-barrier processes).
(b) Model~\cite{Maydanyuk.2011.EPJP} at $k = 1$ (green dash-dotted line 1 for $E_{\rm rad}=100$, blue solid line 2 for $E_{\rm rad}=220$, red dashed line 3 for $E_{\rm rad}=250$).
For $a < 65$ larger duration at $E_{\rm rad} = 220$ in comparison with the duration at $E_{\rm rad} = 100$ is consistent with the previous results for the velocities in Fig.~\ref{fig.6}~(a,~b):
the velocity at $E_{\rm rad} = 100$ is changed more strongly, that causes more strong increase of the duration.
But for larger $a$ one can see similar increase of all durations, that corresponds to similar tendencies of the velocities to unite limit as $a \to 100$.
\label{fig.11}}
\end{figure}

\section{Conclusions}

Process of formation of the universe with its further expansion in the first evolution stage is investigated in the framework of Friedmann-Robertson-Walker metrics on the basis of quantum model, where a new type of matter is introduced, which energy density is dependent on velocity of the expansion.
It is shown that such an improvement of the model forms potential barrier for the flat universe at $k=0$.
Peculiarities of wave function are analyzed in details, which is calculated by fully quantum (non-semiclassical) approach. Resonant influence of the initial and boundary conditions on the barrier penetrability is observed.
In order to analyze dynamics of evolution of the universe in quantum approach, we introduce new operators of velocity of expansion and the function of Hubble.
This basis allows us to study dynamics of evolution of universe in quantum cosmology both in the first stage, and in later times.


\end{document}